\documentclass[11pt]{article}

\usepackage{amssymb}
\usepackage{amsmath}
\usepackage[pdftex]{graphicx}   
\usepackage{color}

\DeclareGraphicsExtensions{.png,.jpg,.eps} 

\newcommand{\atm     }{{\rm ATM}}
\newcommand{\df      }{{\rm DF}}
\newcommand{\rr      }{{\rm RR}}
\newcommand{\call    }{{\rm Call}}
\newcommand{\vega    }{{\rm Vega}}

\newcommand{\rko     }{{\rm RKO}}
\newcommand{\dko     }{{\rm DKO}}
\newcommand{\ot      }{{\rm OT}}
\newcommand{\DOT      }{{\rm DOT}}

\title{\bf Vanna-Volga methods applied to FX derivatives:\\
        from theory to market practice }

\author{Fr\'ed\'eric Bossens\S, Gr\'{e}gory Ray\'ee\dag, Nikos S. Skantzos\P \hspace{1mm} and Griselda Deelstra\ddag \\[3mm]
\S Termeulenstraat 86A, Sint-Genesius-Rode B-1640, Belgium, frederic.bossens@hotmail.com\\[2mm]%
\dag Solvay Brussels School of Economics and Management, Universit\'e Libre de Bruxelles,\\
Avenue FD Roosevelt 50, CP 165, Brussels 1050, Belgium\\[2mm]
\P Tervuursevest 21 bus 104, Leuven B-3001, Belgium, nikos.skantzos@gmail.com\\[2mm]
\ddag Department of Mathematics, Universit\'e Libre de Bruxelles,\\
Boulevard du Triomphe, CP 210, Brussels 1050, Belgium}

\begin{document}

\maketitle

\begin{abstract}
We study Vanna-Volga methods which are used to price first
generation exotic options in the Foreign Exchange market. They are
based on a rescaling of the correction to the Black-Scholes price
through the so-called `probability of survival' and the `expected
first exit time'. Since the methods rely heavily on the
appropriate treatment of market data we also provide a summary of
the relevant conventions. We offer a justification of the core
technique for the case of vanilla options and show how to adapt it
to the pricing of exotic options. Our results are compared to a
large collection of indicative market prices and to more
sophisticated models. Finally we propose a simple calibration
method based on one-touch prices that allows the Vanna-Volga
results to be in line with our pool of market data.
\end{abstract}


\section{Introduction}

\label{sec:intro} The Foreign Exchange (FX) option's market is the
largest and most liquid market of options in the world. Currently,
the various traded products range from simple vanilla options to
first-generation exotics (touch-like options and vanillas with
barriers), second-generation exotics (options with a fixing-date
structure or options with no available closed form value) and
third-generation exotics (hybrid products between different asset
classes). Of all the above the first-generation products receive
the lion's share of the traded volume. This makes it imperative
for any pricing system to provide a fast and accurate
mark-to-market for this family of products. Although using the
Black-Scholes model \cite{blackscholes,merton} it is possible to
derive analytical prices for barrier- and touch -options, this
model is unfortunately based on several unrealistic assumptions
that render the price inaccurate. In particular, the Black-Scholes
model assumes that the foreign/domestic interest rates and the
FX-spot volatility remain constant throughout the lifetime of the
option. This is clearly wrong as these quantities change
continuously, reflecting the traders' view on the future of the
market. Today the Black-Scholes theoretical value (BS TV) is used
only as a reference quotation, to ensure that the involved
counterparties are speaking of the same option.

More realistic models should assume that the foreign/domestic
interest rates and the FX spot volatility follow stochastic
processes that are coupled to the one of the spot. The choice of
the stochastic process depends, among other factors, on empirical
observations. For example, for long-dated options the effect of
the interest rate volatility can become as significant as that of
the FX spot volatility. On the other hand, for short-dated options
(typically less than 1 year), assuming constant interest rates
does not normally lead to significant mispricing. In this article
we will assume constant interest rates throughout.

Stochastic volatility models are unfortunately computationally
demanding and in most cases require a delicate calibration
procedure in order to find the value of parameters that allow the
model reproduce the market dynamics. This has led to alternative
`ad-hoc' pricing techniques that give fast results and are simpler
to implement, although they often miss the rigor of their
stochastic siblings. One such approach is the `Vanna-Volga' (VV)
method that, in a nutshell, consists in adding an analytically
derived correction to the Black-Scholes price of the instrument.
To do that, the method uses a small number of market quotes for
liquid instruments (typically At-The-Money options, Risk Reversal
and Butterfly strategies) and constructs an hedging portfolio
which zeros out the Black-Scholes Vega, Vanna and Volga of the
option. The choice of this set of Greeks is linked to the fact
that they all offer a measure of the option's sensitivity with
respect to the volatility, and therefore the constructed hedging
portfolio aims to take the `smile' effect into account.

The Vanna-Volga method seems to have first appeared in the
literature in \cite{lipton3} where the recipe of adjusting the
Black-Scholes value by the hedging portfolio is applied to
double-no-touch options and in \cite{wystup3} where it is applied
to the pricing of one-touch options in foreign exchange markets.
In \cite{lipton3}, the authors point out its advantages but also
the various pricing inconsistencies that arise from the
non-rigorous nature of the technique. The method was discussed
more thoroughly in \cite{CastagnaMercurio} where it is shown that
it can be used as a smile interpolation tool to obtain a value of
volatility for a given strike while reproducing exactly the market
quoted volatilities. It has been further analyzed in \cite{Fisher}
where a number of corrections are suggested to handle the pricing
inconsistencies. Finally a more rigorous and theoretical
justification is given by \cite{Shkolnikov} where, among other
directions, the method is extended to include interest-rate risk.

A crucial ingredient to the Vanna-Volga method, that is often
overlooked in the literature, is the correct handling of the
market data. In FX markets the precise meaning of the broker
quotes depends on the details of the contract. This can often lead
to treading on thin ice. For instance, there are at least four
different definitions for at-the-money strike (resp.\@, `spot',
`forward', `delta neutral', `50 delta call'). Using the wrong
definition can lead to significant errors in the construction of
the smile surface. Therefore, before we begin to explore the
effectiveness of the Vanna-Volga technique we will briefly present
some of the relevant FX conventions.

The aim of this paper is twofold, namely (i) to describe the
Vanna-Volga method and provide an intuitive justification and (ii)
to compare its resulting prices against prices provided by
renowned FX market makers, and against more sophisticated
stochastic models. We attempt to cover a broad range of market
conditions by extending our comparison tests into two different
`smile' conditions, one with a mild skew and one with a very high
skew. We also describe two variations of the Vanna-Volga method
(used by the market) which tend to give more accurate prices when
the spot is close to a barrier. We finally describe a simple
adjustment procedure that allows the Vanna-Volga method to provide
prices that are in good agreement with the market for a wide range
of exotic options.

To begin with, in section \ref{sec:testbench} we describe the set
of exotic instruments that we will use in our comparisons
throughout. In section \ref{sec:mktdata} we review the market
practice of handling market data. Section \ref{vanvolgmethod} lays
the general ideas underlying the Vanna-Volga adjustment, and
proposes an interpretation of the method in the context of Plain
Vanilla Options. In sections \ref{VVprob} and \ref{VVfet} we
review two common Vanna-Volga variations used to price exotic
options. The main idea behind these variations is to reduce
Vanna-Volga correction through an attenuation factor. The first
one consists in weighting the Vanna-Volga correction by some
function of the survival probability, while the second one is
based on the so-called expected first exit time argument. Since
the Vanna-Volga technique is by no means a self-consistent model,
no-arbitrage constraints must be enforced on top of the method.
This problem is addressed in section \ref{arbtest}. In section
\ref{sec:sensitivies} we investigate the sensitivity of the model
with respect to the accuracy of the input market data. Finally,
Section \ref{numres} is devoted to numerical results. After
defining a measure of the model error in section \ref{typres},
section \ref{sec:Locvol-Stochvol-mktprices} investigates how the
Dupire local vol model \cite{dupire} and the Heston stochastic vol
model \cite{heston} perform in pricing. Section \ref{improv}
suggests a simple adaptation that allows the Vanna-Volga method to
produce prices reasonably in line with those given by renowned FX
platforms. Conclusions of the study are presented in section
\ref{concl}.

\section{Description of first-generation exotics}
\label{sec:testbench}

The family of first-generation exotics can be divided into two
main subcategories: (i) The hedging options which have a strike
and (ii) the treasury options which have no strike and pay a fixed
amount. The validity of both types of options at maturity is
conditioned on whether the FX-spot has remained below/above the
barrier level(s) according to the contract termsheet during the
lifetime of the option.

Barrier options can be further classified as either
\emph{knock-out} options or \emph{knock-in} ones. A knock-out
option ceases to exist when the underlying asset price reaches a
certain barrier level; a knock-in option comes into existence only
when the underlying asset price reaches a barrier level. Following
the no-arbitrage principle, a knock-out plus a knock-in option
(KI) must equal the value of a plain vanilla.

As an example of the first category, we will consider
\emph{up-and-out calls} (UO, also termed Reverse Knock-Out), and
\emph{double-knock-out calls} (DKO). The latter has two knock-out
barriers (one up-and-out barrier above the spot level and one
down-and-out barrier below the spot level). The exact
Black-Scholes price of the UO call can be found in
\cite{hull,reiner1,reiner2}, while a semi-closed form for
double-barrier options is given in \cite{kunitomo} in terms of an
infinite series (most terms of which are shown to fall to zero
very rapidly).

As an example of the second category, we will select
\emph{one-touch} (OT) options paying at maturity one unit amount
of currency if the FX-rate ever reaches a pre-specified level
during the option's life,  and \emph{double-one-touch} (DOT)
options paying at maturity one unit amount of currency if the
FX-rate ever reaches any of two pre-specified barrier levels
(bracketing the FX-spot from below and above). The Black-Scholes
price of the OT option can be found in \cite{rubinstein, wystup},
while the DOT Black-Scholes price is obtained by means of
double-knock-in barriers, namely by going long a double-knock-in
call spread and a double knock-in put spread.

Although these four types of options represent only a very small
fraction of all existing first-generation exotics, most of the
rest can be obtained by combining the above. This allows us to
argue that the results of this study are actually relevant to most
of the existing first-generation exotics.

\section{Handling Market Data}
\label{sec:mktdata}

The most famous defect of the Black-Scholes model is the (wrong)
assumption that the volatility is constant throughout the lifetime
of the option. However, Black-Scholes remains a widespread model
due to its simplicity and tractability. To adapt it to market
reality, if one uses the Black-Scholes formula\footnote{for a
description of our notation, see \ref{app:notation}.}
\begin{eqnarray}
{\rm Call}(\sigma) &=&  {\rm DF}_d(t,T) [FN(  d_1)-KN(  d_2)]\nonumber \\
{\rm Put}(\sigma) &=& -{\rm DF}_d(t,T) [FN(- d_1)-KN(- d_2)]
\label{eq:BS}
\end{eqnarray}
in an inverse fashion, giving as input the option's price and
receiving as output the volatility, one obtains the so-called
`implied volatility'. Plotting the implied volatility as a
function of the strike results typically in a shape that is
commonly termed `smile' (the term `smile' has been kept for
historical reasons, although the shape can be a simple line
instead of a smile-looking parabola). The reasons behind the smile
effect are mainly that the dynamics of the spot process does not
follow a geometric Brownian motion and also that demand for
out-of-the-money puts and calls is high (to be used by traders as
e.g.\@ protection against market crashes) thereby raising the
price, and thus the resulting implied volatility at the edges of
the strike domain.

The smile is commonly used as a test-bench for more elaborate
stochastic models: any acceptable model for the dynamics of the
spot must be able to price vanilla options such that the resulting
implied volatilities match the market-quoted ones. The smile
depends on the particular currency pair and the maturity of the
option. As a consequence, a model that appears suitable for a
certain currency pair, may be erroneous for another.

\subsection{Delta conventions}
\label{sec:delta}

FX derivative markets use, mainly for historical reasons, the
so-called Delta-sticky convention to communicate smile
information: the volatilities are quoted in terms of Delta rather
than strike value. Practically this means that, if the FX spot
rate moves -- all other things being equal -- the curve of implied
volatility vs.\@ Delta will remain unchanged, while the curve of
implied volatility vs.\@ strike will shift. Some argue this brings
more efficiency in the FX derivatives markets. For a discussion on
the appropriateness of the delta-sticky hypothesis we refer the
reader to \cite{Derman}. On the other hand, it makes it necessary
to precisely agree upon the meaning of \emph{Delta}. In general,
Delta represents the derivative of the price of an option with
respect to the spot. In FX markets, the Delta used to quote
volatilities depends on the maturity and the currency pair at
hand. An FX spot $S_t$ quoted as Ccy1Ccy2 implies that 1 unit of
Ccy1 equals $S_t$ units of Ccy2. Some currency pairs, mainly those
with USD as Ccy2, like EURUSD or GBPUSD, use the
\emph{Black-Scholes Delta}, the derivative of the price with
respect to the spot:
\begin{equation}
\Delta_{\rm call} = {\rm DF}_f(t,T)\,N(d_1) \hspace{10mm}
\Delta_{\rm put} = -{\rm DF}_f(t,T)\,N(-d_1) \label{eq:deltas}
\end{equation}
Setting up the corresponding Delta hedge will make one's position
insensitive to small FX spot movements if one is measuring risks
in a USD (domestic) risk-neutral world. Other currency pairs (e.g.
USDJPY) use the \emph{premium included Delta} convention:
\begin{equation}
\tilde{\Delta}_{\rm call} = \frac{K}{S} {\rm DF}_d(t,T)\,N(d_2)
\hspace{10mm} \tilde{\Delta}_{\rm put} = -\frac{K}{S}{\rm
DF}_d(t,T)\,N(-d_2) \label{eq:deltas2}
\end{equation}
The quantities (\ref{eq:deltas}) and (\ref{eq:deltas2}) are
expressed in Ccy1, which is by convention the unit of the quoted
Delta. Taking the example of USDJPY, setting up the corresponding
Delta hedge (\ref{eq:deltas2}) will make one's position
insensitive to small FX spot movements if one is measuring risks
in a USD (foreign) risk-neutral world. Note that (\ref{eq:deltas})
and (\ref{eq:deltas2}) are linked through the option's premium
(\ref{eq:BS}), namely $S(\Delta_{\rm call} - \tilde{\Delta}_{\rm
call}) = {\rm Call}$ and similarly for the put (see
\ref{app:delta} for more details).

With regards to the dependency on maturity, the so-called G11
currency pairs use a spot Delta convention (\ref{eq:deltas}),
(\ref{eq:deltas2}) for short maturities (typically up to 1 year)
while for longer maturities where the interest rate risk becomes
significant, the \emph{forward Delta} (or driftless Delta) is
used, as the derivative of the undiscounted premium with respect
to forward:
\begin{equation}
\begin{array}{ll}
\Delta^F_{\rm call} = N(d_1) & \Delta^F_{\rm put} = -N(-d_1)
\\[3mm]
\tilde{\Delta}^F_{\rm call} = \frac{\displaystyle K}{\displaystyle
S} \frac{{\displaystyle {\rm DF}_d(t,T)}}{{\displaystyle {\rm
DF}_f(t,T)}} N(d_2) & \tilde{\Delta}^F_{\rm put} =
-\frac{\displaystyle K}{\displaystyle S} \frac{{\displaystyle {\rm
DF}_d(t,T)}}{{\displaystyle {\rm DF}_f(t,T)}} N(-d_2)
\end{array}
\label{eq:deltas4}
\end{equation}
where, as before, by tilde we denoted the premium-included
convention. The Deltas in the first row represent the nominals of
the forward contracts to be settled if one is to forward hedge the
Delta risk in a domestic currency while those of the second row
consider a foreign risk neutral world. Other currency pairs
(typically those where interest-rate risks are substantial, even
for short maturities) use the forward Delta convention for all
maturity pillars.

\subsection{At-The-Money Conventions}
\label{sec:ATM}

As in the case of the Delta, the \emph{at-the-money} (ATM)
volatilities quoted by brokers can have various interpretations
depending on currency pairs. The ATM volatility is the value from
the smile curve where the strike is such that the Delta of the
call equals, in absolute value, that of the put (this strike is
termed ATM `straddle' or ATM `delta neutral' ). Solving this
equality yields two possible solutions, depending on whether the
currency pair uses the Black-Scholes Delta or the premium included
Delta convention. The 2 solutions respectively are:
\begin{equation}
 K_{\atm}= F \, \exp \left[ \frac12\sigma_{\atm}^2\tau \right]
\hspace{10mm} \tilde{K}_{\atm}= F \, \exp \left[
-\frac12\sigma_{\atm}^2\tau \right]  \label{atm2}
\end{equation}
Note that these expressions are valid for both \emph{spot} and
\emph{forward} Delta conventions.

\subsection{Smile-related quotes and the broker's Strangle}

Let us assume that a smile surface is available as a function of
the strike $\sigma(K)$. In liquid FX markets some of the most
traded strategies include
\begin{eqnarray}
    {\rm Strangle}(K_c,K_p) &=& \call(K_c,\sigma(K_c)) + {\rm Put}(K_p,\sigma(K_p))
\\
    {\rm Straddle}(K) &=& \call(K,\sigma_\atm) + {\rm Put}(K,\sigma_\atm)
\\
    {\rm Butterfly}(K_p,K,K_c) &=& \frac12\big[{\rm Strangle}(K_c,K_p)-{\rm Straddle}(K)\big]
\end{eqnarray}

Brokers normally quote volatilities instead of the direct prices
of these instruments. These are expressed as functions of
$\Delta$, for instance a volatility at 25$\Delta$-call or put
refers to the volatility at the strikes $K_c,K_p$ that satisfy
$\Delta_{\rm call}(K_c,\sigma(K_c)) = 0.25$ and $\Delta_{\rm
put}(K_p,\sigma(K_p)) = -0.25$ respectively (with the appropriate
Delta conventions, see section \ref{sec:delta}). Typical quotes
for the vols are
\begin{itemize}
\item at-the-money (ATM) volatility: $\sigma_\atm$ \item
25$\Delta$-Risk Reversal (RR) volatility: $\sigma_{\rr 25}$ \item
1-vol-25$\Delta$-Butterfly (BF) volatility: $\sigma_{\rm
BF25(1vol)}$ \item 2-vol-25$\Delta$-Butterfly (BF) volatility:
$\sigma_{\rm BF25(2vol)}$
\end{itemize}
By market convention, the RR vol is interpreted as the difference
between the call and put implied volatilities respectively:
\begin{equation}
\sigma_{\rr 25}=\sigma_{25 \Delta C}-\sigma_{25 \Delta P}
\label{RR}
\end{equation}
where $\sigma_{25\Delta C}=\sigma(K_c)$ and $\sigma_{25\Delta
P}=\sigma(K_p)$.

The 2-vol-25$\Delta$-Butterfly can be interpreted through
\begin{equation}
\sigma_{\rm BF25(2vol)} =\frac{\sigma_{25 \Delta C}+\sigma_{25
\Delta P}}{2}-\sigma_{\atm} \label{BF}
\end{equation}

Associated to the $\sigma_{\rm BF25(2vol)}$ is the
2-vol-25$\Delta$-strangle vol defined through $\sigma_{\rm
STG25(2vol)} = \sigma_{\rm BF25(2vol)} + \sigma_\atm$.

\begin{figure}
\begin{center}
\includegraphics[height=7cm]{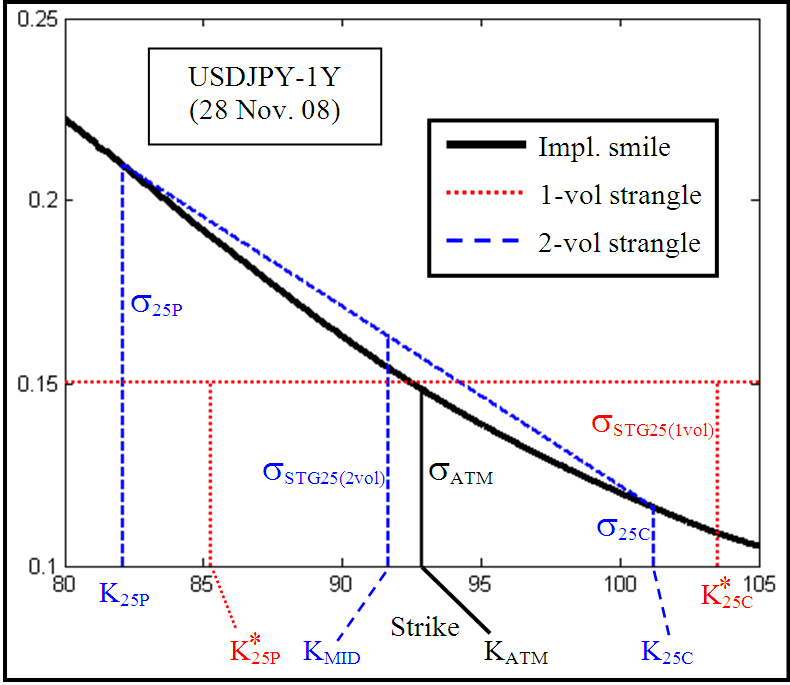}
\includegraphics[height=7cm]{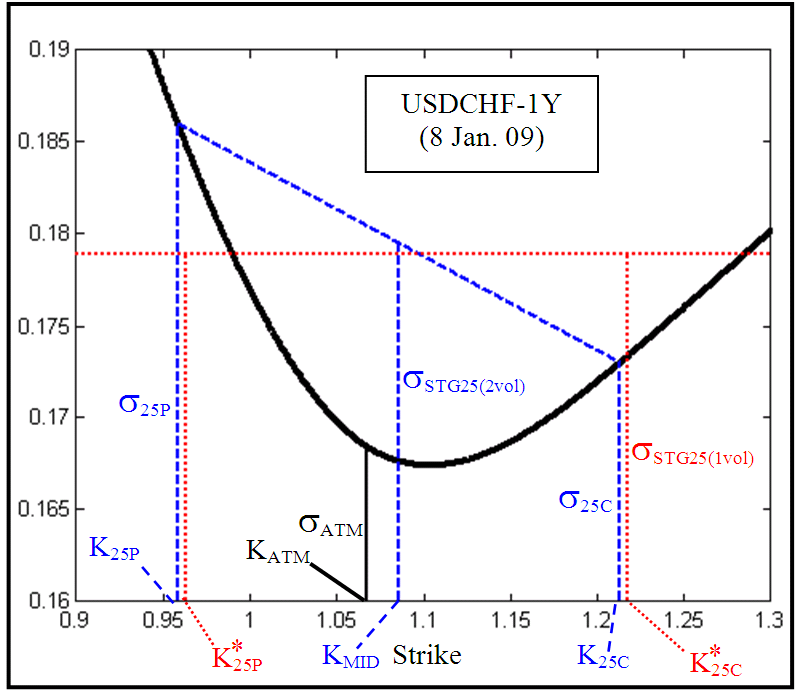}
\caption{Comparison between $\sigma_{\rm STG25(2vol)}$ and
$\sigma_{\rm STG25(1vol)}$, also called `broker strangle' in two
different smile conditions.}\label{fig:brokstrang}
\end{center}
\end{figure}

The $2$-vol-$25$$\Delta$-Butterfly value $\sigma_{\rm BF25(2vol)}$
is in general not directly observable in FX markets. Instead,
brokers usually communicate the $\sigma_{\rm BF25(1vol)}$, using a
\emph{broker's strangle} or \emph{1vol strangle} convention. The
exact interpretation of $\sigma_{\rm BF25(1vol)}$ can be explained
in a few steps:
\begin{itemize}
\item Define $\sigma_{\rm STG25(1vol)}= \sigma_{\rm ATM} +
\sigma_{\rm BF25(1vol)}$. \item  Solve equations
(\ref{eq:deltas}),(\ref{eq:deltas2}) to obtain $K^*_{25C}$ and
 $K^*_{25P}$, the strikes where the Delta of a call is exactly 0.25, and the Delta of
 a put is exactly -0.25 respectively, using the single volatility value $\sigma_{\rm STG25(1vol)}$.
 \item  Provided that the smile curve $\sigma(K)$ is correctly calibrated to the
market, then the quoted value $\sigma_{\rm BF25(1vol)}$ is such
that the following equality holds:
 \begin{eqnarray}
 \call(K^*_{25C},\sigma_{\rm STG25(1vol)}) + {\rm Put}(K^*_{25P},\sigma_{\rm
 STG25(1vol)})  =  \call(K^*_{25C},\sigma(K^*_{25C})) + {\rm Put}(K^*_{25P},\sigma(K^*_{25P})) \label{brokstrang2}
 \end{eqnarray}
\end{itemize}
The difference between $\sigma_{\rm BF25(1vol)}$ and $\sigma_{\rm
BF25(2vol)}$ can be at times confusing. Often for convenience one
sets $\sigma_{\rm BF25(2vol)}=\sigma_{\rm BF25(1vol)}$ as this
greatly simplifies the procedure to build up a smile curve.
However it leads to errors when applied to a steeply skewed
market.  Figure\ref{fig:brokstrang} provides a graphical
interpretation of the quantities $\sigma_{\rm STG25(1vol)}$,
$\sigma_{\rm STG25(2vol)}$, $\sigma_{\rm BF25(1vol)}$ and
$\sigma_{\rm BF25(2vol)}$ in 2 very different market conditions;
the lower panel corresponds to the USDCHF-1Y smile, characterized
by a relatively mild skew, the upper panel corresponding to the
extremely skewed smile of USDJPY-1Y. As a rule of thumb one sets
$\sigma_{\rm BF25(2vol)}=\sigma_{\rm BF25(1vol)}$ when
$\sigma_{\rm RR25}$ is small in absolute value (typically $<1\%$).
When this empirical condition is not met, $\sigma_{\rm
BF25(1vol)}$ and $\sigma_{\rm BF25(2vol)}$ represent actually two
different quantities, and substituting one for the other in  the
context of a smile construction algorithm would yield substantial
errors.

Table \ref{tab:brokstrang} gives more details about the numerical
values used to produce the 2 smiles of Figure
\ref{fig:brokstrang}.

\begin{table}[!h]
\begin{center}
{\begin{tabular}{|c|c|c|}  \cline{2-3} \multicolumn{1}{c|}{} &
USDCHF  & USDJPY \\ \hline date & 8 Jan 09 & 28 Nov 08 \\ \hline
FX spot rate & 1.0902 & 95.47 \\ \hline maturity &
\multicolumn{2}{c|}{1 year} \\ \hline $r_d$ & 1.3\% & 1.74\%   \\
\hline $r_f$ & 2.03\% & 3.74\% \\ \hline $\sigma_{ATM}$ & 16.85\%
& 14.85\% \\ \hline $\sigma_{\rm RR25}$ & -1.3\% & -9.4\% \\
\hline $\sigma_{\rm BF25(2vol)}$ & 1.1\% &
1.45\% \\ \hline $\sigma_{\rm BF25(1vol)}$ &  1.04\% & 0.2\% \\
\hline $ K_{25P}$ / $K_{25C}$ & 0.9586/1.2132 & 82.28/101.25 \\
\hline $ K^*_{25P}$ / $K^*_{25C}$ & 0.9630/1.2179 & 85.24/103.53
\\ \hline
\end{tabular} \caption{Details of market quotes for the two smile curves of
Figure \ref{fig:brokstrang}.\label{tab:brokstrang}} }
\end{center}
\end{table}

Various differences are observed between the 2 smiles. In the
USDCHF case, the values $\sigma_{\rm BF25(2vol)}$ and $\sigma_{\rm
BF25(1vol)}$ are close to each other. Similarly, the strikes used
in the 1vol-25$\Delta$ Strangle are rather close to those attached
to the 2vol-25$\Delta$ Strangle. On the contrary, in the USDJPY
case, large differences are observed between the parameters of the
1vol-25$\Delta$-Strangle and those of the
2vol-25$\Delta$-Strangle.

Unfortunately, there is no direct mapping between $\sigma_{\rm
BF25(1vol)}$ and $\sigma_{\rm BF25(2vol)}$. This is mainly due to
the fact that these two instruments are attached to different
points of the implied volatility curve. The relationship between
$\sigma_{\rm BF25(2vol)}$ and $\sigma_{\rm BF25(1vol)}$ implicitly
depends on the entire smile curve.

In practice however, one may be interested in finding the value of
$\sigma_{\rm BF25(1vol)}$ from an existing smile curve; this can
be achieved using an iterative procedure:
\\
\\
\underline{pseudo-algorithm 1}
\begin{enumerate}
\item  Select an initial guess for $\sigma_{\rm BF25(1vol)}$ \item
compute the corresponding strikes $K^*_{25P}$ and $K^*_{25C}$
\item  assess the validity of equality (\ref{brokstrang2}):
compare the value of the Strangle (i) valued with a unique vol
$\sigma_{\rm BF25(1vol)}$ (ii) valued with 2 implied vol
corresponding to $K^*_{25P}$, respectively $K^*_{25C}$ \item  If
the difference between the two values exceeds some tolerance
level,
                adapt the value $\sigma_{\rm BF25(1vol)}$ and go back to 2.
\end{enumerate}

In case one is given a value of $\sigma_{\rm BF25(1vol)}$ from the
market, and wants to use it to build an implied smile curve, one
may proceed the following way:
\\
\\
\underline{pseudo-algorithm 2}
\begin{enumerate}
\item  Select an initial guess for $\sigma_{\rm BF25(2vol)}$ \item
Construct an implied smile curve using $\sigma_{\rm BF25(2vol)}$
and market value of $\sigma_{\rm RR25}$ \item  Compute the value
of $\sigma_{\rm BF25(1vol)}$ (for instance following guidelines of
pseudo-algorithm 1) \item  Compare $\sigma_{\rm BF25(1vol)}$ you
obtained in 3 to the market-given one. \item  If the difference
between the two values exceeds some tolerance,
                adapt the value $\sigma_{\rm BF25(2vol)}$ and go back to 2.
\end{enumerate}

To close this section on the broker's Strangle issue, let us
clarify another enigmatic concept of FX markets often used by
practitioners, the so-called \emph{Vega-weighted} Strangle quote.
This is in fact an approximation for the value of $\sigma_{\rm
STG25(1vol)}$. To show this, we start from equality
(\ref{brokstrang2}). First we assume $K^*_{25P}=K_{25P}$ and
$K^*_{25C}=K_{25 C}$. Next, we develop both sides in a first order
Taylor expansion in $\sigma$ around $\sigma_{ATM}$. After
canceling repeating terms on the left and right-hand side, we are
left with:
\begin{eqnarray}
&& (\sigma_{\rm STG25(1vol)} - \sigma_{\rm ATM}) \cdot(\mathcal{V}(K_{25P},\sigma_{\rm ATM})+\mathcal{V}(K_{25C},\sigma_{\rm ATM})) \nonumber \\
&& \approx (\sigma_{25\Delta P} - \sigma_{\rm ATM})\cdot
\mathcal{V}(K_{25P},\sigma_{\rm ATM}) +
   (\sigma_{ 25\Delta C} - \sigma_{\rm ATM})\cdot \mathcal{V}(K_{25C},\sigma_{\rm ATM})\nonumber \\ \label{vwstr}
\end{eqnarray}
where $\mathcal{V}(K,\sigma)$ represents the Vega of the option,
namely the sensitivity of the option price $P$ with respect to a
change of the implied volatility: $\mathcal{V}=\frac{\partial
P}{\partial \sigma}$. Solving this for $\sigma_{\rm STG25(1vol)}$
yields:
\begin{equation}
    \sigma_{\rm STG25(1vol)} \approx \frac{\sigma_{\rm 25 \Delta P}\cdot \mathcal{V}(K_{25P},\sigma_{\rm ATM})+ \sigma_{\rm 25 \Delta C}\cdot \mathcal{V}(K_{25C},\sigma_{\rm ATM})}{\mathcal{V}(K_{25P},\sigma_{\rm ATM})+\mathcal{V}(K_{25C},\sigma_{\rm ATM})}
\end{equation}
which corresponds to the average (weighted by Vega) of the call
and put implied volatilities.

Note that according to Castagna et al.\@ \cite{CastagnaMercurio}
practitioners also use the term Vega-weighted butterfly for a
structure where a strangle is bought and an amount of ATM straddle
is sold such that the overall vega of the structure is zero.

\section{The Vanna-Volga Method}

\label{vanvolgmethod}


The Vanna-Volga method consists in adjusting the Black-Scholes TV
by the cost of a portfolio which hedges three main risks
associated to the volatility of the option, the Vega, the Vanna
and the Volga. The Vanna is the sensitivity of the Vega with
respect to a change in the spot FX rate: Vanna = $\frac{\partial
\mathcal{V}}{\partial S}$. Similarly, the Volga is the sensitivity
of the Vega with respect to a change of the implied volatility
$\sigma$: Volga = $\frac{\partial \mathcal{V}}{\partial \sigma}$.
The hedging portfolio will be composed of the following three
strategies:
\begin{eqnarray}
\atm &=& \frac12 {\rm Straddle}(K_\atm) \nonumber \\
\rr &=&  \call(K_c,\sigma(K_c))-{\rm Put}(K_p,\sigma(K_p)) \nonumber \\
\rm{BF} &=& \frac12 {\rm Strangle}(K_c,K_p)- \frac12 {\rm
Straddle}(K_\atm) \label{eq:RR-BF-ATM}
\end{eqnarray}
where $K_\atm$ represents the ATM strike, $K_{c/p}$ the 25-Delta
call/put strikes obtained by solving the equations $\Delta_{\rm
call}(K_c,\sigma_\atm)=\frac14$  and $\Delta_{\rm
put}(K_p,\sigma_\atm)=-\frac14$ and $\sigma(K_{c/p})$ the
corresponding volatilities evaluated from the smile surface.

\subsection{The general framework}

In this section we present the Vanna-Volga methodology.

The simplest formulation \cite{wystup} suggests that the
Vanna-Volga price $X^{\rm VV}$ of an exotic instrument $X$ is
given by
\begin{equation}
X^{\rm VV} = X ^{\rm BS} + \underbrace{\frac{{\rm Vanna}(X)}{{\rm
Vanna}({\rm RR})}}_{w_{\rm RR}} {\rm RR}_{\rm cost} +
\underbrace{\frac{{\rm Volga}(X)}{{\rm Volga}({\rm BF})}}_{w_{\rm
BF}} {\rm BF}_{\rm cost} \label{eq:VVsimple}
\end{equation}
where by $X^{\rm BS}$ we denoted the Black-Scholes price of the
exotic and the Greeks are calculated with ATM volatility. Also,
for any instrument $I$ we define its `smile cost' as the
difference between its price computed with/without including the
smile effect:  ${I}_{\rm cost}={I}_{\rm mkt} - {I}_{\rm BS}$, and
in particular
\begin{eqnarray}
{\rm RR}_{\rm cost} &=&\left[ {\rm Call}(K_c,\sigma(K_c))-{\rm Put}(K_p,\sigma(K_p)) \right] - \left[ {\rm Call}(K_c,\sigma_\atm)-{\rm Put}(K_p,\sigma_\atm) \right] \nonumber \\
{\rm BF}_{\rm cost} &=& \frac12 \left[ {\rm
Call}(K_c,\sigma(K_c))+{\rm Put}(K_p,\sigma(K_p)) \right] -
\frac12 \left[ {\rm Call}(K_c,\sigma_\atm)+{\rm
Put}(K_p,\sigma_\atm) \right]
\end{eqnarray}

The rationale behind (\ref{eq:VVsimple}) is that one can extract
the \emph{smile cost} of an exotic option by measuring the
\emph{smile cost} of a portfolio designed to hedge its Vanna and
Volga risks. The reason why one chooses the strategies BF and RR
to do this is because they are liquid FX instruments and they
carry respectively mainly Volga and Vanna risks. The weighting
factors $w_{RR}$ and $w_{BF}$ in (\ref{eq:VVsimple}) represent
respectively  the amount of RR needed to replicate the option's
Vanna, and the amount of BF needed to replicate the option's
Volga. The above approach ignores the small (but non-zero)
fraction of Volga carried by the RR and the small fraction of
Vanna carried by the BF. It further neglects the cost of hedging
the Vega risk. This has led to a more general formulation of the
Vanna-Volga method \cite{CastagnaMercurio} in which one considers
that within the BS assumptions the exotic option's Vega, Vanna and
Volga can be replicated by the weighted sum of three instruments:
\begin{equation}
\vec{x} = \mathbb{A} \vec{w} \label{VVsystem}
\end{equation}
with
\begin{equation}
\mathbb{A} = \left(\begin{array}{ccc}
{\rm ATM}_{\rm vega} &  {\rm RR}_{\rm vega} & {\rm BF}_{\rm vega} \\
{\rm ATM}_{\rm vanna} &  {\rm RR}_{\rm vanna} & {\rm BF}_{\rm vanna} \\
{\rm ATM}_{\rm volga} &  {\rm RR}_{\rm volga} & {\rm BF}_{\rm
volga}
\end{array}\right)
\hspace{10mm} \vec{w}= \left(\begin{array}{c} w_{\rm ATM} \\
w_{\rm RR} \\ w_{\rm BF}
\end{array}\right)
\hspace{10mm} \vec{x}= \left(\begin{array}{c} X_{\rm vega} \\
X_{\rm vanna} \\ X_{\rm volga}
\end{array}\right)
\label{eq:weights}
\end{equation}
the weightings $\vec{w}$ are to be found by solving the systems of
equations (\ref{VVsystem}).

Given this replication, the Vanna-Volga method adjusts the BS
price of an exotic option by the \emph{smile cost} of the above
weighted sum (note that the ATM smile cost is zero by
construction):
\begin{eqnarray} X^{\rm VV} &=& X
^{\rm BS} + w_{\rm RR} \big({\rm RR}^{\rm mkt}-{\rm RR}^{\rm
BS}\big) +
w_{\rm BF} \big({\rm BF}^{\rm mkt}-{\rm BF}^{\rm BS}\big) \nonumber \\[2mm]
& = & X ^{\rm BS} + \vec{x}^T(\mathbb{A}^T)^{-1}\vec{I} = X ^{\rm
BS} + X_{\rm vega} \, \Omega_{\rm vega}+ X_{\rm vanna} \,
\Omega_{\rm
vanna} + X_{\rm volga} \, \Omega_{\rm volga} \nonumber \\
\label{eq:VV1}
\end{eqnarray}
where
\begin{equation}
\vec{I} = \left(\begin{array}{c}
0 \\
{\rm RR}^{\rm mkt} - {\rm RR}^{\rm BS}\\
{\rm BF}^{\rm mkt} - {\rm BF}^{\rm BS}
\end{array}\right)
\hspace{1cm} \left(\begin{array}{c}
\Omega_{\rm vega} \\
\Omega_{\rm vanna} \\
\Omega_{\rm volga}
\end{array}\right) = (\mathbb{A}^T)^{-1}\vec{I}
\end{equation}
and where the quantities $\Omega_i$ can be interpreted as the
market prices attached to a unit amount of Vega, Vanna and Volga,
respectively. For vanillas this gives a very good approximation of
the market price. For exotics, however, e.g.\@ no-touch options
close to a barrier, the resulting correction typically turns out
to be too large. Following market practice we thus modify
(\ref{eq:VV1}) to
\begin{eqnarray}
X^{\rm VV} & = & X ^{\rm BS} + p_{\rm vanna} X_{\rm vanna} \,
\Omega_{\rm vanna} + p_{\rm volga} X_{\rm volga} \, \Omega_{\rm
volga} \label{eq:VV}
\end{eqnarray}
where we have dropped the Vega contribution which turns out to be
several orders of magnitude smaller than the Vanna and Volga terms
in all practical situations, and where $p_{\rm vanna}$ and $p_{\rm
volga}$ represent attenuation factors which are functions of
either the `survival probability' or the expected `first-exit
time'. We will return to these concepts in section
\ref{sec:variants}.

\subsection{Vanna-Volga as a smile-interpolation method}
\label{sec:taylorexp}

In \cite{CastagnaMercurio}, Castagna and Mercurio show how
Vanna-Volga can be used as a smile interpolation method. They give
an elegant closed-form solution (unique) of system
(\ref{VVsystem}), when $X$ is a European call or put with strike
$K$.\\ In their paper they adjust the Black Scholes price by using
a replicating portfolio composed of a weighted sum of three
vanillas (calls or puts) struck respectively at $K_1$, $K_2$ and
$K_3$, where $K_1<K_2<K_3$. They show that the weights $w_i$
associated to the vanillas struck at $K_i$ such that the resulting
portfolio hedges the Vega, Vanna and Volga risks of the vanilla of
strike $K$ are unique and given by:
\begin{eqnarray}
 w_1(K) & = & \frac{\vega(K)}{\vega(K_1)}\frac{\ln{\frac{K_2}{K}} \ln{\frac{K_3}{K}}} {\ln{\frac{K_2}{K_1}} \ln{\frac{K_3}{K_1}}} \nonumber \\
 w_2(K) & = & \frac{\vega(K)}{\vega(K_2)}\frac{\ln{\frac{K}{K_1}} \ln{\frac{K_3}{K}}} {\ln{\frac{K_2}{K_1}} \ln{\frac{K_3}{K_2}}} \label{weightCastagna} \\
 w_3(K) & = & \frac{\vega(K)}{\vega(K_3)}\frac{\ln{\frac{K}{K_1}} \ln{\frac{K}{K_2}}} {\ln{\frac{K_3}{K_1}} \ln{\frac{K_3}{K_2}}} \nonumber
\end{eqnarray}
The fact that this solution provides an exact interpolation method
is easily verified by noticing that $w_i(K_i)=1$ and $w_i(K_j)=0,
\ i \neq j$.

This solution still holds in the case of a replicating portfolio
composed of ATM, RR and BF instruments as described in section
\ref{vanvolgmethod} by equations (\ref{eq:RR-BF-ATM}). Setting
$K_1=K_p$, $K_2=K_{\atm}$ and $K_3=K_c$, a simple coordinate
transform yields:

\begin{eqnarray}
 w_{\atm}(K) & = & w_1(K) + w_2(K) + w_3(K) \nonumber \\
 w_{\rr}(K) & = & \frac{1}{2} (w_3(K) - w_1(K)) \\
 w_{\rm BF}(K) & = & w_1(K) + w_3(K) \nonumber
\end{eqnarray}

\noindent where the weights $w_{\atm}$, $w_{\rr}$ and $w_{\rm BF}$
are defined by (\ref{VVsystem})-(\ref{eq:weights}).

We now turn back to the elementary Vanna-Volga recipe
(\ref{eq:VVsimple}). Unlike the previously exposed exact solution,
it does not reproduce the market price of RR and BF, a fortiori is
it not an interpolation method for plain vanillas. However, this
approximation possesses the merit of allowing a qualitative
interpretation of the RR and BF correction terms in
(\ref{eq:VVsimple}).

As we will demonstrate, those two terms directly relate to the
slope and convexity of the smile curve. To start with, we
introduce a new smile parametrization variable:
\begin{equation}
Y=\ln\frac{K}{F \cdot \exp{(\frac12 \sigma_{\atm}^2
\tau})}=\ln\frac{K}{K_{\atm}} \nonumber
\end{equation}
Note that the Vega of a Plain Vanilla Option is a symmetric
function of $Y$:
\begin{equation}
{\rm Vega}(Y) ={\rm Vega}(-Y)=S \, e^{-r_f \tau}\, \sqrt{\tau} \,
n\left(\frac{Y}{\sigma \sqrt{\tau}} \right) \nonumber
\end{equation}

\noindent where $n(\cdot)$ denotes the Normal density function.

Let us further assume that the smile curve is a quadratic function
of $Y$:
\begin{equation}
\sigma(Y) = \sigma_\atm + b Y + c Y^2 \label{eq:smile}
\end{equation}
In this way we allow the smile to have a skew (linear term) and a
curvature (quadratic term), while keeping an analytically
tractable expression. We now express Vanna and Volga of Plain
Vanilla Options as functions of $Y$:
\begin{eqnarray}
{\rm Vanna}(Y) &=& {\rm Vega}(Y) \cdot \frac{Y + \sigma^2 \tau}{S
\sigma^2 \tau} \nonumber
\\{\rm Volga}(Y) &=& {\rm Vega}(Y) \cdot \frac{Y^2 + \sigma^2 \tau Y}{\sigma^3 \tau} \label{eq:taylorvvv}
\end{eqnarray}

Working with the plain Black-Scholes Delta (\ref{eq:deltas}) and
the delta-neutral $\atm$ definition and defining
$Y_i=\ln\frac{K_i}{K_{\atm}}$ we have that $Y_{ATM}$ and $Y_{25P}$
and $Y_{25C}$ corresponding respectively to At-The-Money, 25-Delta
Put, and 25-Delta Call solve
\begin{eqnarray*}
&& Y_{ATM} = 0\\
& &    {\rm DF}_f(t,T)\ N\left(\frac{Y_{25P} - \frac12
(\sigma^2_{25P} - \sigma^2_{ATM} ) \tau }{\sigma_{25P}
\sqrt{\tau}}\right) =  \frac14\\
& & {\rm DF}_f(t,T)\ N\left(\frac{-Y_{25C} + \frac12
(\sigma^2_{25C} - \sigma^2_{ATM} ) \tau }{\sigma_{25C}
\sqrt{\tau}}\right) = \frac14
\end{eqnarray*}

Under the assumption that $\sigma_{25\Delta
C}\approx\sigma_{25\Delta P} \approx \sigma_{ATM}$ we find
$Y_{25C}\approx -Y_{25P}$. In this case using equations
(\ref{eq:taylorvvv}) and (\ref{eq:RR-BF-ATM}), the Vanna of the RR
and the Volga of the BF can be expressed as :
\begin{eqnarray}
{\rm Vanna}(RR) &=& {\rm Vanna}(Y_{25C})-{\rm Vanna}(Y_{25P})= 2
\frac{ {\rm Vega}(Y_{25C})Y_{25C} }{S \cdot \sigma_\atm^2 \cdot
\tau}
 \nonumber
\\{\rm Volga}(BF) &=& \frac{{\rm Volga}(Y_{25C})+{\rm Volga}(Y_{25P})}{2}-{\rm Volga}(0)
= \frac{{\rm Vega}(Y_{25C}) Y_{25C}^2}{\sigma_\atm^3
\tau}\nonumber
\\
\label{eq:taylorvv2}
\end{eqnarray}
To calculate $\rm{RR}_{\rm cost}$ and $\rm{BF}_{\rm cost}$ (the
difference between the price calculated with smile, and that
calculated with a constant volatility $\sigma_\atm$), we introduce
the following convenient approximation:
\begin{equation}
{\rm Call}(\sigma(Y))-{\rm Call}(\sigma_\atm) \approx {\rm
Vega}(Y) \cdot \left( \sigma(Y)- \sigma_\atm \right)
\end{equation}
using the above, it is straightforward to show that:
\begin{eqnarray}
{RR}_{\rm cost} &\approx& 2 b \,{\rm Vega}(Y_{25C}) Y_{25C} \nonumber\\
{BF}_{\rm cost} &\approx& c {\rm Vega}(Y_{25C}) \, Y_{25C}^2
\label{eq:RRBFcost}
\end{eqnarray}
Substituting expressions (\ref{eq:taylorvvv}),
(\ref{eq:taylorvv2}) and (\ref{eq:RRBFcost}) in the simple VV
recipe (\ref{eq:VVsimple}) yields the following remarkably simple
result:

\begin{eqnarray}
X^{\rm VV}(Y) &=& X ^{\rm BS}(Y) + \frac{{\rm Vanna}(Y)}{{\rm
Vanna}({\rm RR})} {\rm RR}_{\rm cost} + \frac{{\rm Volga}(Y)}{{\rm
Volga}({\rm BF})} {\rm BF}_{\rm cost} \nonumber
\\ &\approx& X ^{\rm BS} + {\rm Vega(Y)} b Y + {\rm Vega(Y)} c Y^2 + {\rm Vega(Y)}\sigma_\atm^2 \tau \cdot (b+cY) \label{eq:VVsimple2}
\\ &=& X ^{\rm BS}(Y) + \frac{\partial X^{\rm BS}}{\partial \sigma}(Y) \cdot (\sigma(Y) -  \sigma_\atm) + {\rm Vega(Y)}\sigma_\atm^2  \tau \cdot (b+cY) \nonumber
\end{eqnarray}
Despite the presence of a residual term, which vanishes as $\tau
\rightarrow 0$ or $\sigma_\atm \rightarrow 0$, the above
expression shows that the Vanna-Volga price (\ref{eq:VVsimple}) of
a vanilla option can be written as a first-order Taylor expansion
of the BS price around $\sigma_\atm$. Furthermore, as $\frac{{\rm
Vanna}(Y)}{{\rm Vanna}({\rm RR})} {\rm RR}_{\rm cost} \approx {\rm
Vega(Y)} b Y$ and $\frac{{\rm Volga}(Y)}{{\rm Volga}({\rm BF})}
{\rm BF}_{\rm cost} \approx {\rm Vega(Y)} c Y^2 $, the RR term
(coupled to Vanna) accounts for the impact of the linear component
of the smile on the price, while the BF (coupled to Volga)
accounts for the impact of the quadratic component of the smile on
the price.

\section{Market-adapted variations of Vanna-Volga }
\label{sec:variants}

In this section we describe two empirical ways of adjusting the
weights $\big(p_{\rm vanna},p_{\rm volga}\big)$ in (\ref{eq:VV}).
We will focus our attention on knock-out options, although the
Vanna-Volga approach can be readily generalized to options
containing knock-in barriers, as those can always be decomposed
into two knock-out (or vanilla) ones (through the no-arbitrage
relation knock-in = vanilla -- knock-out).

To justify the need for the correction factors to (\ref{eq:VV}) we
argue as follows: As the knock-out barrier level $B$ of an option
is gradually moved toward the spot level $S_t$, the BS price of a
KO option must be a monotonically decreasing function, converging
to zero exactly at $B=S_t$. Since the Vanna-Volga method is a
simple rule-of-thumb and not a rigorous model, there is no
guarantee that this will be satisfied. We thus have to impose it
through the attenuations factors $p_{\rm vanna}$ and $p_{\rm
volga}$. Note that for barrier values close to the spot, the Vanna
and the Volga behave differently: the Vanna becomes large while,
on the contrary, the Volga becomes small. Hence we seek
attenuation factors of the form:
\begin{equation}
p_{\rm vanna} = a \, \gamma  \hspace{1cm} p_{\rm volga} = b + c \,
\gamma \label{eq:pvannavolga}
\end{equation}
where $\gamma\in[0,1]$ represents some measure of the barrier(s)
vicinity to the spot with the features
\begin{eqnarray}
\gamma=0 &{\rm for}& S_t\to B  \label{eq:limit1}\\
\gamma=1 &{\rm for}& |S_t-B|\gg 0  \label{eq:limit2}
\end{eqnarray}
i.e.\@ the limiting cases refer to the regions where the spot is
close versus away from the barrier level. Before moving to more
specific definitions of $\gamma$, let us introduce some
restrictions on $p_{\rm vanna}$ and $p_{\rm volga}$:
\begin{equation}
\lim_{\gamma \rightarrow 1} p_{\rm vanna} = 1 \hspace{1cm}
\lim_{\gamma \rightarrow 1} p_{\rm volga} = 1
\end{equation}
The above conditions ensure that when the barrier is far from the
spot, implying that hitting the barrier becomes very unlikely, the
Vanna-Volga algorithm boils down to its simplest form
(\ref{eq:VVsimple}) which is a good approximation to the price of
a vanilla option using the market quoted volatility. We therefore
amend the expressions (\ref{eq:pvannavolga}) into continuous,
piecewise linear functions:
\begin{equation}
p_{\rm vanna} = \left\{
\begin{array}{ll}
 a \gamma  & \gamma \leq \gamma^*  \\
 a \gamma^* \frac{1-\gamma}{1-\gamma^*} + \frac{\gamma-\gamma^*}{1-\gamma^*} & \gamma > \gamma^*
\end{array} \right.
\hspace{1cm} p_{\rm volga} = \left\{
\begin{array}{ll}
b + c \, \gamma & \gamma \leq \gamma^* \\
(b + c \, \gamma^*)\frac{1-\gamma}{1-\gamma^*} +
\frac{\gamma-\gamma^\star}{1-\gamma^*} & \gamma > \gamma^*
\end{array} \right.  \label{eq:pasympt}
\end{equation}
where $\gamma^*$ is a transition threshold chosen close to 1. Note
that the amendment (\ref{eq:pasympt}) is justified only in the
case of options that degenerate into plain vanilla instruments in
the region where the barriers are away from the spot. However, in
the case of treasury options that do not have a strike (e.g.\@
OT), there is no smile effect in the region where the barriers are
away from the spot as these options pay a fixed amount and their
fair value is provided by the BS TV. In this case, no amendment is
necessary as both Vanna and Volga go to zero.

We now proceed to specify practical $\gamma$ candidates, namely
the {\em survival probability} and the expected {\em first exit
time} (FET). In what follows, the corresponding Vanna-Volga prices
will be denoted by ${\rm VV}_{\rm surv}$ and ${\rm VV}_{\rm fet}$
respectively.

\subsection{Survival probability}
\label{VVprob}

The survival probability  $p_{\rm surv}\in[0,1]$ refers to the
probability that the spot does not touch one or more barrier
levels before the expiry of the option. Here we need to
distinguish whether the spot process is simulated through the
domestic or the foreign risk-neutral measures:
\begin{eqnarray}
{\rm domestic:} && dS_t = S_t\,(r_d-r_f)\,dt+ S_t\,\sigma\,dW_t \label{eq:wiener1}\\
{\rm foreign:} &&  dS_t = S_t\,(r_d-r_f+\sigma^2)\,dt+
S_t\,\sigma\,dW_t \label{eq:wiener2}
\end{eqnarray}
where $W_t$ is a Wiener process. One notices that the quanto drift
adjustment will obviously have an impact in the value of the
survival probability. Then, for e.g.\@ a single barrier option we
have
\begin{eqnarray}
{\rm domestic:} & & p_{\rm surv}^d = \texttt{E}^d[ 1_{S_{t'}<B, t<t'<T}] = {\rm NT}^d(B) / {\rm DF}_d(t,T) \\
{\rm foreign:} &        & p_{\rm surv}^f = \texttt{E}^f[
1_{S_{t'}<B, t<t'<T}] = {\rm NT}^f(B) / {\rm DF}_f(t,T)
\end{eqnarray}
where ${\rm NT}^{d/f}(B)$ is the value of a no-touch option in the
domestic/foreign measure, $\texttt{E}^{d/f}$ is the risk neutral
expectation in the domestic/foreign market respectively, and $1_a$
is the indicator function for the event ``a''. Similarly, for
options with two barriers the survival probability is given
through the undiscounted value of a double-no-touch option.
Explicit formulas for no-touch and double-no-touch options can be
found in \cite{wystup}.

The survival probability clearly satisfies the required features
(\ref{eq:limit1}), (\ref{eq:limit2}). To respect domestic/foreign
symmetries we further define $\gamma_{\rm surv}= \frac12(p_{\rm
surv}^d + p_{\rm surv}^f)$.

\subsection{First exit time}
\label{VVfet}

The first exit time is the minimum between: (i) the time in the
future when the spot is expected to exit a barrier zone before
maturity, and (ii) maturity, if the spot has not hit any of the
barrier levels up to maturity. That is, if we denote the FET by
$u(S_t,t)$ then $u(S_t,t) = \min\{\phi,\tau\}$ where
$\phi=\inf\{\ell\in [0,\infty) | S_{t+\ell} > H\ {\rm or}\
S_{t+\ell}<L\}$ where $L<S_{t}<H$ define the barrier levels, $S_t$
the spot of today and $\tau$ the time to maturity (expressed in
years). This quantity also has the desirable feature that it
becomes small near a barrier and can therefore be used to rescale
the two correction terms in (\ref{eq:VV}).

Let us give some definitions. For a geometric Brownian motion spot
process of constant volatility $\sigma$ and drift $\mu$, the
cumulative probability of the spot hitting a barrier between $t^*$
and $t'$ ($t<t^*<T$, $t'>t^*$) denoted by $C(S,t^*,t')$ obeys a
backward Kolmogorov equation \cite{wilmott} (in fact $C(S,t^*,t')$
can be thought of as the undiscounted price of a DOT option):
\begin{equation}
\mathcal{F}\, C = 0 \hspace{10mm} \mathcal{F}\equiv \frac{\partial
}{\partial t^*} + \frac12\sigma^2 S^2 \frac{\partial^2}{\partial
S^2} + \mu S \frac{\partial }{\partial S} \label{eq:FETPDE}
\end{equation}
with boundary conditions $C(L,t^*,t')=C(H,t^*,t')=1$ and
$C(S,t',t')=0$ assuming that there are no
window-barriers\footnote{In a window-barrier option, the barrier
is activated at a time greater than the selling time of the option
and deactivates before the maturity of the option.}.
Now suppose that at some time $t^*>t$, we are standing at $S$, and
no barrier was hit so far, the expected FET (measured from $t$) is
then by definition:
\begin{equation}
\overline{u(S,t^*)} = t^*-t + \int_{t^*}^{T} (t'-t^*)
\frac{\partial C}{ \partial t'}\,dt' + \int_{T}^{\infty} (T-t^*)
\frac{\partial C}{ \partial t'}\,dt'
\end{equation}
while integration by parts gives
\begin{equation}
\overline{u(S,t^*)} = t^*-t + \int_{t^*}^T \left( 1 - C(S,t^*,t')
\right) \,dt'
\end{equation}
and finally taking derivative with respect to $t^*$, and first and
second derivatives with respect to $S$ and integrating
(\ref{eq:FETPDE}) from $t^*$ to $T$ results in:
\begin{equation}
\frac{\partial \overline{u}}{\partial t^*} + \frac12\sigma^2 S^2
\frac{\partial^2 \overline{u}}{\partial S^2} + \mu S
\frac{\partial \overline{u}}{\partial S} = 0 \Leftrightarrow
\mathcal{F}\, \overline{u} = 0 \label{eq:FETPDE2}
\end{equation}
note that this is slightly different from the expression in
\cite{wilmott}, where FET is measured from $t^*$. Equation
(\ref{eq:FETPDE2}) is solved backwards in time from $t^*=T$ to
$t^*=t$, starting from the terminal condition $\overline{u(S,T)} =
\tau$ and boundary conditions
$\overline{u(L,t^*)}=\overline{u(H,t^*)}=t^*-t$. In case of a
single barrier option we use the same PDE with either $H\gg S_t$
or $L\ll S_t$.

As for the case of the survival probability we solve the PDE
(\ref{eq:FETPDE2}) in both the domestic and foreign risk-neutral
cases which implies that we set as parameters of (\ref{eq:FETPDE})
\begin{eqnarray}
{\rm domestic:} & & \sigma=\sigma_{\rm ATM}, \ \mu = r_d-r_f\\
{\rm foreign:} & & \sigma=\sigma_{\rm ATM}, \ \mu = r_d-r_f +
\sigma^2
\end{eqnarray}
where $r_d$ and $r_f$ correspond to the Black-Scholes domestic and
foreign interest rates. Let us denote the solution of the above
PDE as $\lambda^d$ and $\lambda^f$ respectively. Finally we define
$\gamma_{\rm fet} = \frac12 \frac{(\lambda^d+\lambda^f)}{\tau}$.
Note that we have divided by the time to maturity in order to have
a dimensionless quantity with $\gamma_{\rm fet} \in[0,1]$.

\subsection{Qualitative differences between $\gamma_{\rm surv}$ and $\gamma_{\rm fet}$}
\label{sec:tradeoff}

Although $\gamma_{\rm surv}$ and $\gamma_{\rm fet}$ possess
similar asymptotic behavior (converging to 0 for options
infinitely close to knocking-out, converging to 1 for an option
infinitely far from knocking-out), they represent different
quantities, and can differ substantially in intermediate
situations. To support this assertion, we show in Figure
\ref{fig:FETvsPROB} plots of $\gamma_{\rm surv}$ and $\gamma_{\rm
fet}$ as a function of the barrier level, in a single-barrier and
in a double-barrier case. While in the single barrier case the
shapes of the two curves look similar, their discrepancy is more
pronounced in the double-barrier case where the upper barrier is
kept constant, and the lower barrier is progressively moved away
from the spot level. For barrier levels close to the spot, there
is a {\em plateau} effect in the case of $\gamma_{\rm surv}$ which
stays at zero, while $\gamma_{\rm fet}$ seems to increase
linearly. This can be explained intuitively: moving the barrier
level in the close vicinity of the spot will not prevent the spot
from knocking out at some point before maturity (hence
$\gamma_{\rm surv} \approx 0$). But although the knocking event is
almost certain, the expected time at which it occurs directly
depends on the barrier-spot distance.

This discussion should emphasize the importance of a careful
choice between the two $\gamma$ candidates, especially when it
comes to pricing double-barrier options.

There is no agreed consensus regarding which of $\gamma_{\rm
surv},\gamma_{\rm fet}$ is a better candidate for $\gamma$ in
({\ref{eq:pvannavolga}). Based on empirical observations, it is
suggested in e.g.\@ \cite{Stein} that one uses $\gamma_{\rm surv}$
with $a=1$ and $b=c=0.5$. Other market beliefs however favor using
$\gamma_{\rm fet}$ with $a=c=1$ and $b=0$. In \cite{Wystup2}, the
absence of mathematical justification for these choices is
highlighted, and other adjustment possibilities are suggested,
depending on the type of option at hand. In section \ref{numres}
we will discuss a more systematic procedure that can allow one to
calibrate the Vanna-Volga model and draw some conclusions
regarding the choice of pricer.

\begin{figure}
\begin{center}
\includegraphics[height=8cm]{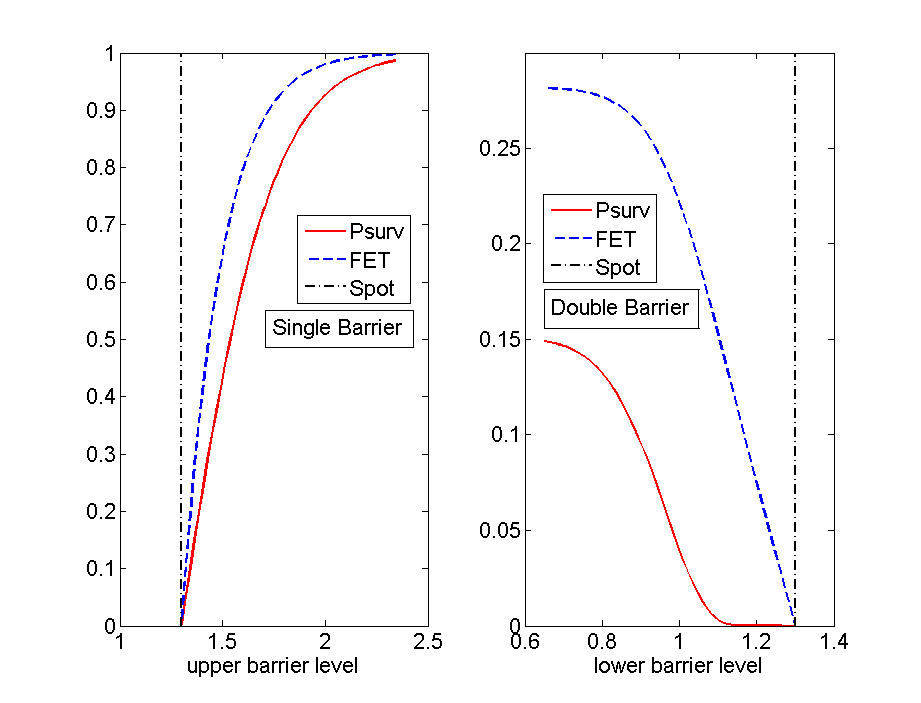}
\caption{Comparison between $\gamma_{\rm surv}$ and $\gamma_{\rm
fet}$ plotted against barrier level, in a single barrier case
(left panel) and a double barrier case (right panel). Used Market
Data: $S$=1.3, $\tau$=1.3, $r_d$=5\%, $r_f$=3\%
$\sigma_\atm$=20\%}
 \label{fig:FETvsPROB}
\end{center}
\end{figure}

\subsection{Arbitrage tests}

\label{arbtest} As the Vanna-Volga method is not built on a solid
bedrock but is only a practical rule-of-thumb, there is no
guarantee that it will be arbitrage free. Therefore as part of the
pricer one should implement a testing procedure that ensures a few
basic no-arbitrage rules for barrier options (with or without
strike): For example, (i) the value of a vanilla option must not
be negative, (ii) the value of a single/double knock-out barrier
option must not be greater than the value of the corresponding
vanilla, (iii) the value of a double-knock-out barrier option must
not be more expensive than either of the values of the
corresponding single knock-outs, (iv) the value of a window
single/double knock-out barrier option must be smaller than that
of the corresponding vanilla and greater than the corresponding
american single/double knock-out. For knock-in options, the
corresponding no-arbitrage tests can be derived from the
replication relations: (a) for single barriers, KI(B) = VAN --
KO(B), where B represents the barrier of the option, and (b) for
double-barriers, KIKO(KIB,KOB) = KO(KIB) -- DKO(KIB,KOB) where KIB
and KOB represent the knock-in and knock-out barrier respectively.

For touch or no-touch options, the above no-arbitrage principles
are similar. One-touch options can be decomposed into a discounted
cash amount and no-touch options: OT(B) = DF - NT(B) and similarly
for double-one-touch options.

Based on these principles a testing procedure can be devised that
amends possible arbitrage inconsistencies. We begin by using
replication relations to decompose the option into its constituent
parts if needed. This leaves us with vanillas and knock-out
options for which we calculate the BSTV and the Vanna-Volga
correction. On the resulting prices we then impose
\begin{equation}
{\rm VAN} = \max({\rm VAN},0) \hspace{10mm} {\rm KO} = \max({\rm
KO},0) \label{eq:positivity}
\end{equation}
to ensure condition (i) above. We then proceed with imposing
conditions (ii)-(iv):
\begin{equation}
{\rm KO} = \min({\rm KO},{\rm VAN}) \hspace{5mm} {\rm WKO} =
\min({\rm WKO},{\rm VAN}) \hspace{5mm} {\rm WKO} = \max({\rm
WKO},{\rm KO}) \label{eq:KOvsVAN}
\end{equation}
while for double-knock-out options we have in addition
\begin{equation}
{\rm DKO} = \min({\rm DKO},{\rm KO(1)}) \hspace{10mm} {\rm DKO} =
\min({\rm DKO},{\rm KO(2)}) \hspace{10mm} \label{eq:DKOcond}
\end{equation}
where KO(1) and KO(2) represent the corresponding single knock-out
options.

Note that both in the case of a double-knock-out and in that of a
window-knock-out we need to create a single-knock-out instrument
and launch a no-arbitrage testing on it as well.

As an example, let us consider a window knock-in knock-out option.
Having an `in' barrier this option will be decomposed to a
difference between a window knock-out and a window double
knock-out. For the former, we will create the corresponding KO
option while for the latter the corresponding DKO. In addition, we
will also need the plain vanilla instrument. We will then price
the KO and DKO separately using the Vanna-Volga pricer, ensure
that the resulting value of each of these is positive (equation
(\ref{eq:positivity})), impose condition (iii) (equation
(\ref{eq:DKOcond})) to ensure no-arbitrage on the DKO and
condition (iv)  (equation (\ref{eq:KOvsVAN})) to ensure that the
barrier options are not more expensive than the plain vanilla.

\subsection{Sensitivity to market data}
\label{sec:sensitivies}

As the FX derivatives market is rife with complex conventions it
can be the case that pricing errors stemming from wrong input data
have a greater impact than errors stemming from assuming wrong
smile dynamics. This warrants discussion concerning the
sensitivity of FX models with respect to market data. Already from
(\ref{eq:VVsimple}) we can anticipate that the Vanna-Volga price
is sensitive to the values of $\sigma_{\rm RR25}$ and $\sigma_{\rm
BF25(2vol)}$.  To emphasize this dependency we will consider the
following two sensitivities:
\begin{equation}
\Lambda_{\rm RR} = \frac{\rm d\,Price}{{\rm d}\,\sigma_{ \rm
RR25}} \hspace{10mm} \Lambda_{\rm BF} = \frac{\rm d\,Price}{{\rm
d}\,\sigma_{\rm BF25(2vol)}} \label{eq:mktsens}
\end{equation}
which measure the change in the Vanna-Volga price given a change
in the input market data. In our tests we have used the
Vanna-Volga `survival probability' for a series of barrier levels
of a OT option. Similar considerations follow by using the FET
variant. The results are shown in Figure \ref{fig:DPDRR} where on
top of the two sensitivities we superimposed the Vanna and the
Volga of the option.

\begin{figure}
\begin{center}
\includegraphics[height=8cm]{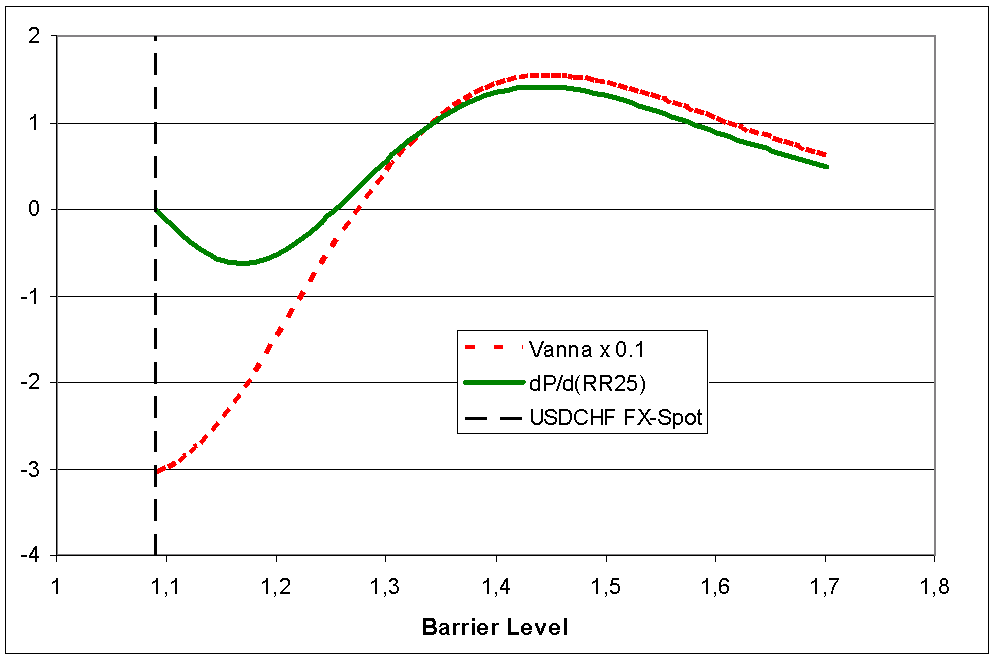}
\includegraphics[height=8cm]{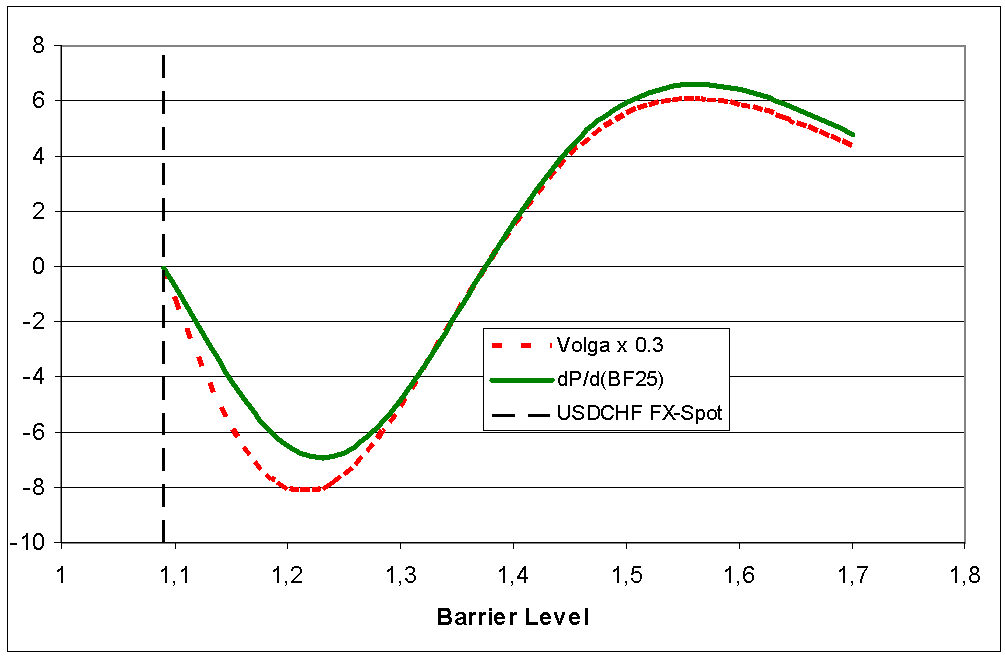}
\caption{Sensitivity of the Vanna-Volga price with respect to
input market data for a OT option. Top: Comparison between the
Vanna (BSTV) and $\Lambda_{\rm RR}$. Bottom: Comparison between
the Volga (BSTV) and $\Lambda_{\rm BF}$. We see that the two
Greeks provide a good approximation of the two model
sensitivities.} \label{fig:DPDRR}
\end{center}
\end{figure}

We notice that the two sensitivities can deviate significantly
away from zero. This highlights the importance of using accurate
and well-interpreted market quotes. For instance, in the 1-year
USDCHF OT with the touch-level at 1.55 (BSTV price is $\approx
4\%$), an error of $0.5\%$ in the value of $\sigma_{\rm
BF25(2vol)}$ would induce a price shift of $3\%$. This is all but
negligible! Thus a careful adjustment of the market data quotes is
sometimes as important as the model selection.

We also see that the Volga provides an excellent estimate of the
model's sensitivity to a change in the Butterfly values.
Similarly, Vanna  provides a good estimate of the model's
sensitivity to a change in the Risk Reversal values --but only as
long as the barrier level is sufficiently away from the spot. This
disagreement in the region close to the spot is linked to the fact
that in the Vanna-Volga recipe of section \ref{VVprob} we adjusted
the Vanna contribution by the survival probability which becomes
very small close to the barrier.

Figure \ref{fig:DPDRR} implies that for all practical purposes one
should be on guard for high BS values of Vanna and/or Volga which
indicate that the pricer is sensitively dependent on the accuracy
of the market data.

\section{Numerical results}
\label{numres}

In order to assess the ability of the Vanna-Volga family of models
(\ref{eq:VV}) to provide market prices, we compared them to a
large collection of market indicative quotes. By \emph{indicative}
we mean that the prices we collected come from trading platforms
of three major FX-option market-makers, queried without
effectively proceeding to an actual trade. It is likely that the
models behind these prices do not necessarily follow demand-supply
dynamics and that the providers use an analytic pricing method
similar to the Vanna-Volga we present here.

Our pool of market prices comprises of 3-month and 1-year options
in USDCHF and USDJPY, the former currency pair typically
characterized by small $\sigma_{\rr}$ values, while the latter by
large ones. In this way we expect to span a broad range of market
conditions. For each of the four maturity/currency pair
combinations we select four instrument types, representative of
the first generation exotics family: Reverse-Knock-Out call (RKO),
One-Touch (OT), Double-Knock-Out call (DKO), and Double-One-Touch
(DOT). In the case of single barrier options (RKO and OT), 8
barrier levels are adjusted, mapping to probabilities of touching
the barrier that range from 10\% to 90\%. In the case of the RKO
call, the strike is set \emph{At-The-Money-Spot}. In the case of
two-barrier options (DKO and DOT), since it is practically
impossible to fully span the space of the two barriers we selected
the following subspace: (i) we fix the lower barrier level in such
a way that it has a 10\% chance of being hit, then select 5 upper
barrier levels such that the overall hitting probabilities (of any
of the 2 barriers) range approximately from 15\% to 85\%. (ii) We
repeat the same procedure with a fixed upper barrier level, and 5
adjusted lower barrier levels.

In summary, our set of data consists of the cross product
$\mathcal{F}$ of the sets
\begin{eqnarray}
{\rm currency\ pair:} && A=\{ {\rm USDJPY}, {\rm USDCHF}\}  \nonumber \\
{\rm maturity\ period:} && B=\{ 3m, 1y \}  \nonumber  \\
{\rm option\ type:} && C=\{ \rko, \ot, \dko, \DOT \} \nonumber  \\
{\rm barrier\ value:} && D=\{ B_1, \ldots, B_n\} \label{eq:pool}
\end{eqnarray}
where $n=10$ for double-barrier options and $n=8$ for single
barrier ones.

In order to maintain coherence, each of the two data sets were
collected in a half-day period (in Nov. 2008 for USDJPY, in Jan.
2009 for USDCHF).

Thus in total our experiments are run over the set of models
\begin{equation}
{\rm models:} \hspace{5mm} E = \{{\rm VV}_{\rm surv}, {\rm
VV}_{\rm fet}\}
\end{equation}

\subsection{Definition of the model error}

\label{typres} In order to focus on the smile-related part of the
price of an exotic option, let us define for each instrument
$i\in\mathcal{F}$ from our pool of data (\ref{eq:pool}) the `Model
Smile Value' (MODSV) and the `Market Smile Value' (MKTSV) as the
difference between the price and its Black-Scholes Theoretical
Value (BSTV):
\begin{eqnarray}
{\rm MODSV }_i^k  &=&  {\rm Model\ Price}_i^k - {\rm BSTV}_i^k \hspace{10mm} k=1,\ldots,N_{\rm mod}\nonumber \\
{\rm MKTSV }_i^k  &=&  {\rm Market\ Price}_i^k - {\rm BSTV}_i^k
\hspace{10mm} k=1,\ldots,N_{\rm mkt} \nonumber
\end{eqnarray}
(where market prices  are taken as the average between bid and ask
prices) and where $N_{\rm mod}=4$ is the number of models we are
using and $N_{\rm mkt}=3$ the number of FX market makers where the
data is collected from. Let us also define the average, minimum
and maximum of the market smile value:
\begin{eqnarray} && \overline{{\rm MKTSV}_i } =
\frac{1}{N_{\rm mkt}}\sum_{k\leq N_{\rm mkt}} {\rm MKTSV}_i^k
\nonumber \\ &&  {\rm min}_i  = \min_{k\in N_{\rm mkt}} {\rm
MKTSV}_i^k \ , \hspace{5mm} {\rm max}_i = \max_{k\in N_{\rm mkt}}
{\rm MKTSV}_i^k
\end{eqnarray}
We now introduce an error measure quantifying the ability of a
model to describe market prices. This function is defined as a
quadratic sum over the pricing error :
\begin{equation}
\varepsilon_k=\sum_{i\in\mathcal{F}} \left( \frac{{\rm
MODSV}_i^k-\overline{{\rm MKTSV}_i }} { {\rm max}_i- {\rm min}_i}
\right)^2 \label{eq:errfunc}
\end{equation}
The error is weighted by the inverse of the market spread, defined
as the difference between the maximum and the minimum mid market
price for a given instrument. This setup is designed (i) to yield
a dimensionless error measure that can be compared across currency
pairs and the type of options, (ii) to link the error penalty to
the market coherence: a pricing error on an instrument which is
priced very similarly by the 3 market providers will be penalized
more heavily than the same pricing error where market participants
exhibit large pricing differences among themselves. Note also that
the error is defined as the deviation from the average market
price.

\subsection{Shortcomings of common stochastic models in pricing exotic options}
\label{sec:Locvol-Stochvol-mktprices}

Before trying to calibrate the Vanna-Volga weighting factors
$p_{\rm vanna}$ and $p_{\rm volga}$, we investigate how the Dupire
local vol \cite{dupire} and the Heston stochastic vol
\cite{heston} models perform in pricing our set of selected exotic
instruments (for a discussion on the pricing of barrier
instruments under various model frameworks, see for example
\cite{lipton1,lipton2,lipton3}). In order to obtain a fast and
reliable calibration for Heston, the price of call options is
numerically computed through the characteristic function
\cite{albrecher,jackel}, and Fourier inversion methods. To price
exotic options, Heston dynamics is simulated by Monte Carlo, using
a Quadratic-Exponential discretization scheme \cite{andersen}.

Figure \ref{fig:HestonDupire} shows the MODSV of a 1-year OT
options in USDCHF (lower panel) and USDJPY (upper panel), as the
barrier moves away from the spot level ($S_t=95.47$ for USDJPY and
$S_t=1.0902$ for USDCHF). At first inspection, none of the models
gives satisfactory results.


\begin{figure}
\begin{center}
\includegraphics[height=7cm]{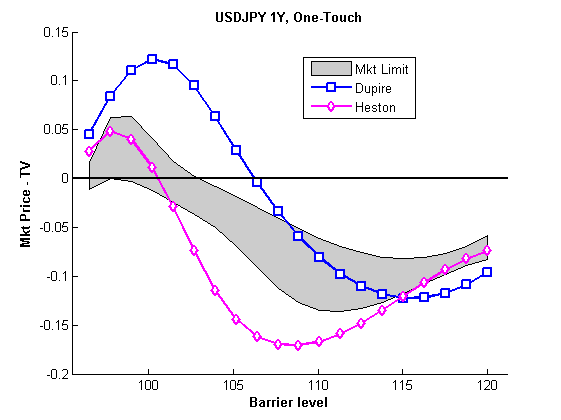}
\includegraphics[height=7cm]{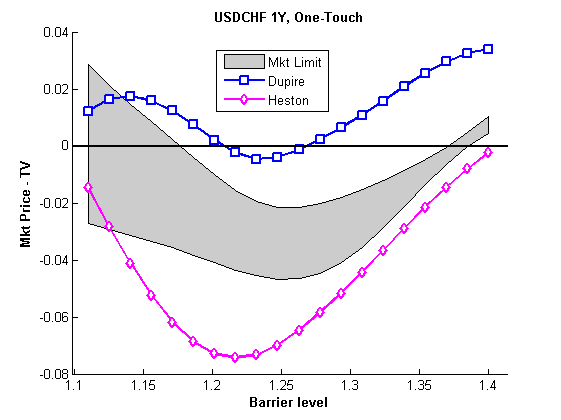}
\caption{Smile value vs.\@ barrier level; comparison of the
various models for OT 1-year options in  USDCHF (bottom) and
USDJPY (top). Market limits are indicated with black solid lines.}
\label{fig:HestonDupire}
\end{center}
\end{figure}

\begin{table}[!h]
\begin{center}
{\begin{tabular}{|c|c|c|c|} \cline{3-4} \multicolumn{2}{c|}{} &
USDCHF  & USDJPY \\ \hline
 & RKO & Heston & Dupire \\ \cline{2-4}
 & OT  & Heston & Heston \\ \cline{2-4}
 & DKO(Up)  & Dupire & Dupire \\ \cline{2-4}
 1-Year & DKO(Down)  & Heston & Heston \\ \cline{2-4}
 & DOT(Up)  & Heston & Dupire \\ \cline{2-4}
 & DOT(Down)  & Heston & Dupire \\ \cline{2-4}
 & \bf{global}  & \bf{Heston ($\varepsilon=62$)} & \bf{Dupire ($\varepsilon=96$)} \\ \cline{2-4} \hline
 & RKO & Heston & Dupire \\ \cline{2-4}
 & OT  & Heston & Dupire \\ \cline{2-4}
 & DKO(Up)  & Heston & Dupire \\ \cline{2-4}
 3-Month & DKO(Down)  & Heston & Heston \\ \cline{2-4}
 & DOT(Up)  & Heston & Dupire \\ \cline{2-4}
 & DOT(Down)  & Heston & Heston \\ \cline{2-4}
 & \bf{global}  & \bf{Heston ($\varepsilon=65$)} & \bf{Dupire  ($\varepsilon=73$)} \\ \cline{2-4} \hline
\end{tabular}\caption{Heston stochastic vol Vs. Dupire local vol in pricing 1st
generation exotics.\label{tab:HestonDupire}}}
\end{center}
\end{table}

Using the error measure defined above, we now try to formalize the
impressions given by our rough inspection of Figure
\ref{fig:HestonDupire}. For each combination of the instruments in
(\ref{eq:pool}) we determine which of Dupire local vol or Heston
stochastic vol gives better market prices. The outcome of this
comparison is given in the Table \ref{tab:HestonDupire}.

This table suggests that --in a simplified world where exotic
option prices derive either from Dupire local vol or from heston
stochastic vol dynamics-- an FX market characterized by a mild
skew (USDCHF) exhibits mainly a stochastic volatility behavior,
and that FX markets characterized by a dominantly skewed implied
volatility (USDJPY) exhibit a stronger local volatility component.
This confirms that calibrating a stochastic model to the vanilla
market is by no mean a guarantee that exotic options will be
priced correctly \cite{Schoutens}, as the vanilla market carries
no information about the smile dynamics.

In reality the market dynamics could be better approximated by a
hybrid volatility model that contains both some stochastic vol
dynamics and some local vol one. This model will be quite rich but
the calibration can be expected to be considerably hard, given
that it tries to mix two very different smile dynamics, namely an
`absolute' local-vol one with a `relative' stochastic vol one. For
a discussion of such a model, we refer the reader to
\cite{lipton2}.

At this stage one has the option to either go for the complex
hybrid model or for the more heuristic alternative method like the
Vanna-Volga. In this paper we present the latter.

\subsection{Vanna-Volga calibration}

\label{improv}

The purpose of this section is to provide a more systematic
approach in selecting the coefficients $a$, $b$ and $c$ in
(\ref{eq:pvannavolga}) and thus the factors $p_{\rm vanna}$ and
$p_{\rm volga}$.

We first determine the optimal values of coefficients $a$, $b$ and
$c$ in the sense of the least error (\ref{eq:errfunc}), where the
sum extends to all instruments and to the two maturities (e.g. a
single error function per currency pair). This problem can readily
be solved using standard linear regression tools, as $a$, $b$ and
$c$ appear linearly in the VV correction term, but most standard
solver algorithms would as well do the job. This optimization
problem is solved four times in total, for USDCHF with
$\gamma_{\rm surv}$ and $\gamma_{\rm fet}$, and for USDJPY with
$\gamma_{\rm surv}$ and $\gamma_{\rm fet}$. Let us point out that
such a calibration is of course out of the question in a real
trading environment: collecting such an amount of market data each
time a recalibration is deemed necessary would be way too
time-consuming. Our purpose is simply to determine some limiting
cases, to be used as benchmarks for the results of a more
practical calibration procedure discussed later. Table
\ref{tab:optsol} presents these optimal solutions, indicating the
minimum error value, along with the value of the optimal
coefficients $a$, $b$ and $c$.

\begin{table}[!h]
\begin{center}
 {\begin{tabular}{|c|c|c|} \cline{2-3}
\multicolumn{1}{c|}{} & USDCHF  & USDJPY \\
\hline
$\gamma_{\rm surv}$  & $\bf{\varepsilon=19.7}$ &  $\bf{\varepsilon=15.6}$ \\
                     & $a=0.54$, $b=0.29$, $c=0.14$ & $a=0.74$, $b=0.7$, $c=0.05$ \\
\hline
$\gamma_{\rm fet}$  & $\bf{\varepsilon=18.2}$ &  $\bf{\varepsilon=14.7}$ \\
                     & $a=0.49$, $b=0.35$, $c=0.01$ & $a=0.54$, $b=0.17$, $c=0.52$ \\
\hline
\end{tabular}\caption{Overall pricing error, calibration on entire market price
set.\label{tab:optsol}}}
\end{center}
\end{table}

Comparing the above error numbers to those of Table
\ref{tab:HestonDupire}, it seems possible that the Vanna-Volga
models have the potential to outperform the Dupire or Heston
models.

We now discuss a more practical calibration approach, where the
minimization is performed on OT prices only. The question we try
to answer is: `Can we calibrate a VV model on OT market prices,
and use this model to price other first generation exotic products
?'. Performing this calibration with 3 parameters to optimize will
certainly improve the fitting of OT prices, but at the expense of
destroying the fitting quality for the other instruments (in the
same way that performing a high-order linear regression on a set
of data points, will produce a perfect match on the data points
and large oscillations elsewhere). This is confirmed by the
results of Table \ref{tab:optsol2}, showing how the error (on the
entire instrument set) increases with respect to the error of
Table \ref{tab:optsol} when the optimization is performed on the
OT subset only.

\begin{table}[!h]
\begin{center}
{\begin{tabular}{|c|c|c|} \cline{2-3}
\multicolumn{1}{c|}{} & USDCHF  & USDJPY \\
\hline
$\gamma_{\rm surv}$  & $\bf{\varepsilon=44.6}$ &  $\bf{\varepsilon=26.8}$ \\
\hline
$\gamma_{\rm fet}$  & $\bf{\varepsilon=47.2}$ &  $\bf{\varepsilon=85}$ \\
\hline
\end{tabular}\caption{Overall pricing error, calibration on OT prices
only.\label{tab:optsol2}} }
\end{center}
\end{table}

For robustness reasons, it is thus desirable to reduce the space
of free parameters in the optimization process. We consider the
following two constrained optimization setups: (i) $a=c$, $b=0$
and (ii) $b=c=0.5\cdot a$, which are re-scaled versions of the
market practices described in section \ref{sec:tradeoff}. Needless
to say that the number of possible configurations here are limited
only by one's imagination. Our choice is dictated mainly by
simplicity, namely we have chosen to keep a single degree of
freedom. The results are presented in Table \ref{tab:optsol3}
where we compare four possible configurations measured over all
instruments and maturity periods for our two currency pairs.

\begin{table}[!h]
\begin{center}
{\begin{tabular}{|c|c|c|c|c|} \cline{3-5}
\multicolumn{2}{c|}{} & USDCHF  & USDJPY & Total error\\
\hline
configuration 1 & $\gamma_{\rm surv}$  & $\varepsilon=21.8$ &  $\varepsilon=28.4$ & $\varepsilon$=50.2\\
 & $b=c=0.5\cdot a$ & $a=0.43$ &  $a=0.72$ & \\
\hline
configuration 2 & $\gamma_{\rm fet}$  & $\varepsilon=21.2$ &  $\varepsilon=26$ & $\varepsilon$=47.2\\
 & $b=c=0.5\cdot a$ & $a=0.39$ &  $a=0.63$ & \\
\hline
configuration 3 & $\gamma_{\rm surv}$  & $\varepsilon=32.2$ &  $\varepsilon=72.1$ &  $\varepsilon$=104.3\\
& $a=c$, $b=0$ & $a=0.51$ &  $a=0.69$ & \\
\hline
configuration 4 & $\gamma_{\rm fet}$  & $\varepsilon=24.3$ &  $\varepsilon=19.4$ & $\varepsilon$=43.7\\
& $a=c$, $b=0$ & $a=0.42$ &  $a=0.60$ & \\
\hline
\end{tabular}\caption{Overall pricing error, constrained calibration on OT
prices only.\label{tab:optsol3}}}
\end{center}
\end{table}

As there is no sound mathematical (or economical) argument to
prefer one configuration over another, we therefore choose the
least-error configuration, namely configuration $n^o$4. One
additional argument in favor of $\gamma_{\rm fet}$ is that it
accommodates window-barrier options without further adjustment.
This is not the case of $\gamma_{\rm surv}$ where some re-scaling
should be used to account for the start date of the barrier (when
the barrier start date is very close to the option maturity, the
path-dependent character vanishes and the full VV correction
applies i.e. $p_{\rm vanna}=p_{\rm volga}=1$ even for small
$\gamma_{\rm surv}$ values).

\begin{figure}
\begin{center}
\includegraphics[height=6cm]{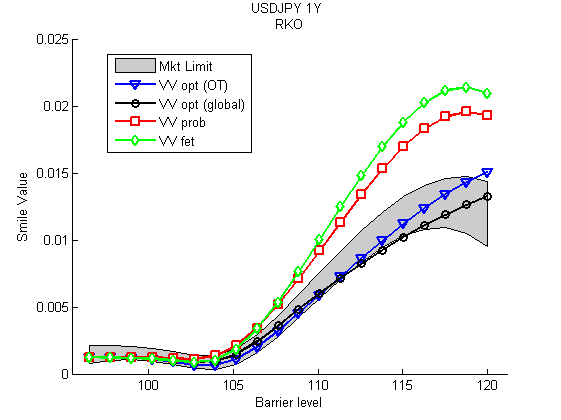}\includegraphics[height=6cm]{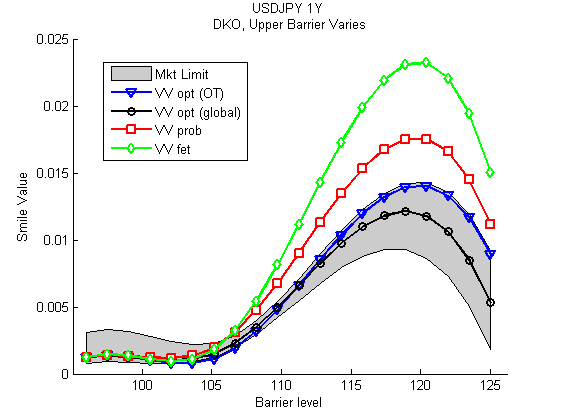}
\includegraphics[height=6cm]{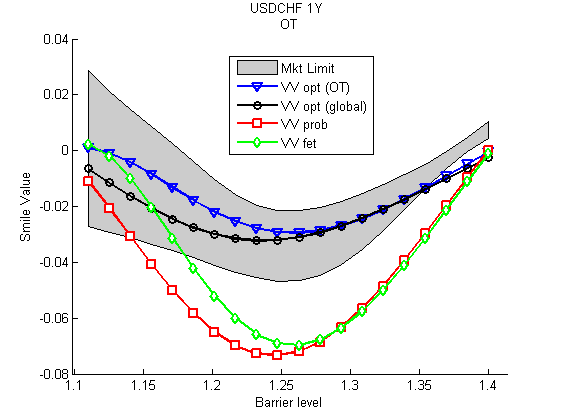}\includegraphics[height=6cm]{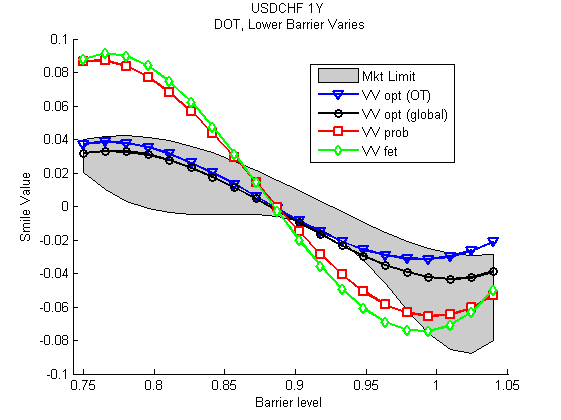}
\caption{\small Results from calibrating the Vanna-Volga method on
(i) all instruments of our data pool (marked as `VV opt
(global))', (ii) one-touch options only (marked as `VV opt (OT))'.
The results of the two calibrations do not differ significantly
while the latter is naturally more convenient from a practical
perspective. The shaded areas correspond to the region within
which market makers provide their indicative mid price. For
comparison we also show the non-calibrated Vanna-Volga methods
based on the `survival probability' and the `first exit time'.}
\label{fig:mainres}
\end{center}
\end{figure}

In Figure \ref{fig:mainres} we show results from the calibration
of the Vanna-Volga method. It is based on minimizing the error
(\ref{eq:errfunc}) of (i) all instruments of the data pool and
while having all coefficients $a,b,c$ of $\gamma_{\rm fet}$ free
and (ii) of one-touch options only and with configuration $n^o$4
(thus, we have chosen $\gamma_{\rm fet}$ with $a=c$, $b=0$). We
see that in general calibration (i) performs better in the sense
that it falls well within the shaded area that corresponds to the
limits of the market price as provided by the FX market makers.
This is not surprising as this calibration is meant to be the most
general and flexible. However this is clearly an impractical
calibration procedure. On the contrary, the calibration method
(ii) that is based on quotes from a single exotic instrument has
practical advantages and appears in good agreement with that of
(i). Finally note that these pictures are representative of our
results in general.

\section{Conclusion}
\label{concl}

The Vanna-Volga method is a popular pricing tool for FX exotic
options. It is appealing to both traders, due to its clear
interpretation as a hedging tool, and to quantitative analysts,
due to its simplicity, ease of implementation and computational
efficiency. In its simplest form, the Vanna-Volga recipe assumes
that smile effects can be incorporated to the price of an exotic
option by inspecting the effect of the smile on vanilla options.
Although this recipe, outlined in (\ref{eq:VVsimple}), turns out
to give often uncomfortably large values, there certainly is a
silver lining there. This has led market practitioners to consider
several ways to adapt the Vanna-Volga method. In this article we
have reviewed some commonly used adaptations based on rescaling
the Vanna-Volga correction by a function of either the `survival
probability' or the `first exit time'. These variations provide
prices that are more in line with the indicative ones given by
market makers.

We have attempted to improve the Vanna-Volga method further by
adjusting the various rescaling factors that are involved. This
optimization is based on simple data analysis of one-touch options
that are obtained from renowned FX platforms. It involves a single
optimization variable and as a result we find that for a wide
range of exotic options, maturity periods and currency pairs it
leads to prices that agree well with the market mid-price.

The FX derivatives community, perhaps more than any other asset
class, lives on a complex structure of quote conventions.
Naturally, a wrong interpretation of the input market data cannot
lead to the correct results. To this end, we have presented some
relevant FX conventions regarding smile quotes and we have tested
the robustness of the Vanna-Volga method against the input data.
It appears that the values of Vanna and Volga provide a good
indication of the VV price sensitivity to a change in smile input
parameters.

\newpage

\appendix

\section{Definitions of notation used}
\label{app:notation}

\begin{table}[!h]
\begin{center}
{\begin{tabular}{|| l || l ||}
        \hline
                    $t$ & date of today\\
                    $T$ & maturity date\\
                    $\tau=(T-t)/365$ & time to expiry (expressed in years)\\
             $S_t$ & spot today \\
             $K$ & strike \\
             $ r_{f/d}(t)$ & foreign/domestic interest rates \\
             $ \sigma $ & volatility of the FX-spot  \\
             ${\rm DF}_{f/d}(t,T)=\exp[-r_{f/d} \tau]$  &  foreign/domestic discount factor\\
             $ F = S_t {\rm DF}_f(t,T)/{\rm DF}_d(t,T) $  &  forward price \\
             $d_{1} = \frac{\ln\frac{F}{K} + \frac12\sigma^2\tau}{\sigma \sqrt{\tau}}$  & \\
             $d_{2} = \frac{\ln\frac{F}{K} - \frac12\sigma^2\tau}{\sigma \sqrt{\tau}}$  & \\
             $N(z) = \int_{-\infty}^z dx \frac{1}{\sqrt{2\pi}} e^{-\frac12x^2}$ & cumulative normal  \\
        \hline
        \end{tabular}\caption{List of abbreviations.}}
        \end{center}
\end{table}

\section{Premium-included Delta}
\label{app:delta}

For correctly calculating the Delta of an option it is important
to identify which of the currencies represents the risky asset and
which one represents the riskless payment currency.

Let us consider a generic spot quotation in terms Ccy1-Ccy2
representing the amount of Ccy2 per unit of Ccy1. If the
(conventional) premium currency is Ccy2 (e.g.\@ USD in EURUSD)
then by convention the `risky' asset is Ccy1 (EUR in this case)
while Ccy2 refers to the risk-free one. In this case the standard
Black-Scholes theory applies and the Delta expressed in Ccy1 is
found by a simple differentiation of (\ref{eq:BS}): $\Delta_{\rm
BS} = {\rm DF}_f(t,T)\,N(d_1)$. This represents an amount of Ccy1
to sell if one is long a Call.

If, however, the premium currency is Ccy1 (e.g.\@ USD in USDJPY)
then Ccy2 is considered as the risky asset while Ccy1 the
risk-free one. In this case, the value of the Delta is $\Delta =
S_t\Delta_{\rm BS} -\call_t $, where $\call_t$ is the premium in
units of Ccy2 while $\Delta$ and $\Delta_{\rm BS}$ are expressed
in their `natural' currencies; Ccy2 and Ccy1, respectively (for
lightening notations, we omit the time index $t$ in $\Delta$ and
$\Delta_{\rm BS}$). In this case $\Delta$ represents an amount of
Ccy2 to buy. This relation can be seen by the following argument.
First note that the Black-Scholes vanilla price of a call option
is
\begin{equation}
\call_t  = \df_d(t,T)\,\mathbb{E}^d\big[\max(S_T-K,0)\big]
\label{eq:domcall}
\end{equation}
where the index `$d$' implies that we are referring to the
domestic risk-neutral measure, i.e.\@ we take the domestic
money-market (MM) unit $1/\df_d(t,T)$ as numeraire. If we now wish
to express (\ref{eq:domcall}) into a measure where the numeraire
is the foreign money-market account then
\begin{equation}
\call_t  = \df_d(t,T)\,\mathbb{E}^d\big[\max(S_T-K,0)\big] =
\df_d(t,T)\,\mathbb{E}^f\big[\frac{dQ^d}{dQ^f}(T)\,\max(S_T-K,0)\big]
\end{equation}
where we introduced the Radon-Nikodym derivative (see for example
\cite{BrigoMercurio,Shreve})
\begin{equation}
\frac{dQ^d}{dQ^f}(T) = \frac{\df_f(t,T)}{\df_d(t,T)}\ \frac{ S_t}
{S_T}
\end{equation}
This equality allows us to derive the foreign-domestic parity
relation
\begin{equation}
\call_t = \df_d(t,T) \, \mathbb{E}^d\big[\max(S_T-K,0)\big] =
\df_f(t,T) \, S_t\, K\,
\mathbb{E}^f\left[\max(\frac1K-\frac{1}{S_T},0)\right]
\end{equation}
where both sides are expressed in units of Ccy2 (for a unit
nominal amount in Ccy1). The above foreign/domestic relation
illustrates the fact that in FX any derivative contract can be
regarded either from a domestic or from a foreign standpoint.
However the contract value is unique. On the contrary, the Delta
of the option depends on the adopted perspective. In `domestic'
vs.\@ `foreign' worlds we have respectively
\begin{equation}
\Delta_{\rm BS} = \frac{\partial \call_t}{\partial S_t}
\hspace{10mm} \Delta = - \frac{\partial
\frac{\call_t}{S_t}}{\partial \frac{1}{S_t}}
\end{equation}
where the first equation is expressed in units of Ccy1 (to sell)
while the second in units of Ccy2 (to buy). Setting up a Delta
hedged portfolio (at time $t$) in the `foreign' world implies that
at any instant of time $t'>t$, where $t$ represents today, the
portfolio in Ccy1
\begin{equation}
\Pi_{t'} = \frac{\call_{t'}}{S_{t'}} + \frac{\Delta}{S_{t'}}
\end{equation}
will be insensitive to variations of the spot $S_t$. From
$\partial \Pi_{t'} /\partial S_{t'}|_{t'=t} = 0$ we then find
\begin{equation}
\Delta = S_t\,\Delta_{\rm BS} - \call_{t}
\end{equation}
Note that FX convention dictates that the $\Delta$ is always
quoted in units of Ccy1 (regardless of the currency to which the
premium is paid), hence to obtain the relation mentioned in
section \ref{sec:delta} we simply take $\Delta \to \tilde{\Delta}
= \Delta / S_t$. Table \ref{tab:delta} provides a
\emph{vis-\`a-vis} of the various quantities under the two
perspectives for an option in USDJPY with the Spot $S_t$ defined
as  the amount of JPY per USD.

\begin{table}[h]
\begin{center}
{\begin{tabular}{|l||c|c|} \cline{2-3}
\multicolumn{1}{c|}{} & USD world & JPY world\\
\hline
Local MM unit  & 1 USD &  1 JPY \\

Risky asset  & JPY &  USD \\

Contract value in local MM units & $\call_t / S_t$ & $\call_t$ \\

Risky asset in local MM units & $1/S_t$ & $S_t$\\[3mm]

$\Delta$ hedge: amount of risky asset to short & $\frac{\partial
\frac{\call_t}{S_t}}{\partial
\frac{1}{S_t}}=\call_t-S_t\Delta_{\rm BS}$ (JPY)
& $\frac{\partial \call}{\partial S_t}=\Delta_{\rm BS}$ (USD) \\[3mm]

Amount of USD to short & $\tilde{\Delta}=-\frac{\partial
\frac{\call}{S_t}}{\partial
\frac{1}{S_t}}\frac{1}{S_t}=\Delta_{\rm BS}-\frac{1}{S_t}\call$
(USD)
& $\frac{\partial \call}{\partial S_t}=\Delta_{\rm BS}$ (USD)\\
\hline
\end{tabular}\caption{Delta hedge calculation, domestic versus foreign
world.\label{tab:delta}}}
\end{center}
\end{table}

\end{document}